\documentclass[aps,pra,amssymb,amsfonts,floatfix,showpacs]{revtex4}

\usepackage{graphicx,psfrag}
\usepackage{dcolumn}
\usepackage{bm}
\usepackage{mathbbol}
\begin{document}

\title{Advantages of Randomization in Coherent Quantum Dynamical Control}

\author{Lea F. Santos}
\email{lsantos2@yu.edu}
\affiliation{\mbox{Department of Physics, 
Yeshiva University, 245 Lexington Ave, New York, NY 10016, USA}}
\author{Lorenza Viola}
\email{Lorenza.Viola@Dartmouth.edu}
\affiliation{\mbox{Department of Physics and Astronomy, 
Dartmouth College, 6127 Wilder Laboratory, Hanover, NH 03755, USA}}

\date{\today}

\begin{abstract}
Control scenarios have been identified where the use of randomized design 
may substantially improve the performance of dynamical decoupling methods
[L. F. Santos and L. Viola, Phys. Rev. Lett. {\bf 97}, 150501 (2006)].
Here, by focusing on the suppression of internal unwanted interactions in 
closed quantum systems, we review and further elaborate on the advantages of randomization at long evolution times. By way of illustration, special emphasis is devoted to isolated Heisenberg-coupled chains of spin-1/2 particles. In particular, for nearest-neighbor interactions, two types of decoupling cycles are contrasted: inefficient averaging, whereby the number of control actions increases exponentially with the system size, and efficient averaging associated to a fixed-size control group.  The latter allows for analytical and numerical studies of efficient decoupling schemes created by exploiting and merging together randomization and deterministic strategies, such as symmetrization, concatenation, and cyclic permutations.  Notably, sequences capable to remove interactions up to third order are explicitly constructed.  The consequences of faulty controls are also analyzed.
\end{abstract}


\maketitle

\tableofcontents

\section{Introduction}

Dynamical decoupling (DD) provides a versatile control-theoretic setting for manipulating the dynamics of closed as well as open quantum systems. 
DD schemes operate by subjecting the system of interest to suitable sequences of external control operations, with the purpose of removing or modifying unwanted contributions to the underlying Hamiltonian.  DD methods have a long history in high-resolution nuclear magnetic resonance 
(NMR)~\cite{HaeberlenBook,ErnstBook,LevittBook}, where coherent averaging  ideas have been pioneered in the context of removing undesired phase evolution~\cite{Hahn-Echo,CP-Echo} and dipolar interactions~\cite{Waugh} in spin systems.  More recently, DD has emerged as a promising strategy
toward achieving scalable quantum information processing (QIP), thanks to its potential for protecting logical quantum states against always-on qubit-qubit interactions and for suppressing environmental decoherence. The latter possibility was explicitly demonstrated in~\cite{Viola1998} -- where suppression of decoherence via a sequence of very fast (so-called {\em bang-bang}) control actions is described for a single qubit interacting with a bosonic environment -- and it was soon incorporated within a general dynamical symmetrization framework~\cite{Viola1,ZanardiSym0} -- whereby the DD operations are drawn from a discrete control group so to effectively project out components with unintended symmetry.  Since then, DD has become the subject of intense theoretical and experimental investigations.  On the theoretical side, some notable advances include: the construction of bounded-strength Eulerian~\cite{Viola2003Euler} and concatenated DD 
protocols~\cite{Khodjasteh2004,Khodjasteh2005}, as well as efficient combinatorial schemes for multipartite systems~\cite{Jones,Stoll,Leung02,Wocjan}; the identification of optimized control sequences capable to ensure exact high-order cancellation of pure dephasing in a single qubit~\cite{Uhrig2007,UDD}; proposed applications within specific (notably, solid-state) scalable quantum computing architectures~\cite{ss}; quantitative investigations of DD schemes for compensating specific decoherence mechanisms, such as magnetic
state decoherence in atomic systems~\cite{Search}, 1/$f$ noise in superconducting devices~\cite{Shiokawa2002,Gutmann2003,Falci2004,Faoro2004,Gutmann2004},
and hyperfine- as well as phonon-induced decoherence in quantum dots~\cite{Sousa,Sham,Witzel,Wen01,Wen03,Tom08}. Within experimental QIP, DD techniques have been successfully applied to decoherence control in a single-photon polarization interferometer~\cite{Berglund2000}; have found extensive applications in liquid-state NMR QIP~\cite{Cory-Overview}, including in conjunction with error-correcting codes \cite{BoulantDec}; have inspired charge-based~\cite{Nakamura2002} and flux-based~\cite{Chiorescu} echo experiments in superconducting qubits; and are being scrutinized for further applications in solid-state systems such as nuclear quadrupole qubits~\cite{Fraval} and fullerene qubits \cite{Morton2005}.

Even if, in the absence of any control constraint and under appropriate mathematical assumptions, DD techniques may guarantee the exact elimination of all the undesired coupling, a main limitation is the fact that, in general, such an exact averaging is practically impossible.
Residual errors arising from imperfect averaging accumulate in time and eventually result in loss of fidelity.  In order to slow down error accumulation, randomization may be incorporated into DD design, as proposed in \cite{Viola2005Random,Violacdc05}.  In a sense, this is reminiscent of compensation schemes, which are routinely used in NMR spectroscopy to reduce the effects of known errors introduced by non-ideal control~\cite{HaeberlenBook,LevittBook}. Essentially, randomization aims at compensating for imperfect averaging by 
enforcing {\em probabilistic error build-up} at long times, the overall coherent DD action being retained provided the applied control history is appropriately recorded \cite{Viola2005Random,Violacdc05}.  
Beside long-time averaging, analytical errors bounds in~\cite{Viola2005Random,Violacdc05} identified two other scenarios where randomized protocols would be expected to perform better than their deterministic counterparts: first, whenever the basic decoupling cycle
requires a large number of control operations; and, second, when the interactions to be removed are uncertain, for instance unpredictably fluctuating in time. This prompted a series of quantitative studies to validate the advantages of randomization in the context of DD 
\cite{Kern2005,Santos2005,Kern2006,Santos2006,Viola2006Snow,Kern2007}, and ultimately added quantum control to the list of problems benefiting from stochasticity; a list which already includes diverse phenomena ranging from the possibility to maximize weak signals by stochastic
resonance~\cite{GammaSR}, the idea that chaos may stabilize quantum algorithms~\cite{Prosen}, and yet, more recently, the fact that randomization may be used to benchmark noisy quantum gates in QIP
~\cite{Knill2007}.

It is the purpose of this work to both comprehensively review and further analyze the advantages of randomized coherent-control methods at long evolution times.  We focus on the representative case of suppression of internal interactions in a time-independent spin-$1/2$ Heisenberg-coupled system.  In order to pinpoint the origin of the advantages coming from randomization, 
two scenarios are considered: averaging over an inefficient DD group, whose size increases exponentially with the number of spins; and averaging over a small, fixed-size group. The first case allows for a numerical comparison between deterministic and randomized schemes as the group size increases. The second lends itself to a detailed analytical study of various high-level deterministic schemes, which employ symmetrization, concatenation, and cyclic permutations -- eventually leading to the identification of best-performing deterministic DD scheme. The incorporation and analysis of different randomized strategies to further boost the performance of the resulting deterministic schemes is then carried out numerically.

Ultimately, the key idea for efficient averaging at long times is frequent scrambling of the order of the applied DD operations, so that residual errors do not get a chance to rapidly accumulate in time.  While this idea is at the heart of randomized methods, a natural question is why randomization would have to be invoked in the first place: What prevents one from finding an optimal deterministic sequence for a specific system and a particular final time?  The problem lies in the fact that, due to the
rapidly growing number of possible control trajectories associated with different sequences as the system size increases, combined with the strong dependence of protocol performance on the final time, such search is typically intractable in practice. We illustrate this point by developing a numerical algorithm to obtain the best DD sequence under some constraints imposed to the controls. Although very efficient, the resulting sequence is still outperformed by a considerably simpler randomized protocol. We therefore advocate that, for long evolution times,
a much less demanding and yet very efficient decoupling approach consists in cleverly combining good deterministic strategies with randomization.

The content of the paper is organized as follows. In Sec.~II, the theoretical framework of DD is briefly recalled, as well as the  performance metric and interaction frames relevant to the subsequent discussion. Sec.~III describes the deterministic and randomized protocols to be compared and present analytical lower bounds for their expected 
performance under ideal control assumptions. Sec.~IV discusses the models to be studied and highlights the main control requirements. Focus is given to systems with nearest-neighbor couplings and to the ability to selectively address individual spins. In Sec.~V, we compare how the performance of deterministic and randomized schemes depend on the size of the DD group. Schemes involving a large degree of parallelism emerge as 
best performers, consistent with intuition. The core of the paper is contained in Sec.~VI. There, we present analytical studies for deterministic protocols; introduce a new deterministic sequence; compare 
numerically deterministic and randomized schemes, as well as different  venues for including randomization; discuss the results obtained with different systems; and propose an algorithm to search for efficient DD sequences. A comparison of deterministic vs. randomized DD protocols
would not be complete without the inclusion of some dominant control errors, which is done in Sec.~VII. Conclusions and discussions 
are provided in Sec.~VIII. Technical considerations are left for the Appendix.

\section{Dynamical Decoupling Framework}

\subsection{Control setting}

As mentioned, DD methods have long been applied in  
NMR spectroscopy~\cite{CP-Echo,ErnstBook,HaeberlenBook,LevittBook}, 
where the aim is to modify the nuclear spin Hamiltonian to suppress or scale selected internal interactions. More recently, DD has been revisited in the light of quantum control theory, by also explicitly addressing, in particular, the removal of interactions between the system of interest and
the surrounding environment \cite{Viola1998,Viola1}.
In both cases, the basic idea consists in adding an appropriate time-dependent control field $H_c(t)$ to the Hamiltonian $H_0$ of the relevant
target system. In the physical (Schr\"odinger) frame, the evolution operator under the total Hamiltonian $H(t)=H_0+H_c(t)$ becomes 
$ U(t) ={\cal T} \exp [- i\int _{0}^t H(u) du]$, where $\hbar$ is set equal to 1 and ${\cal T}$ denotes time ordering. Most commonly, the analysis of DD methods is performed in a logical frame (also known as ``toggling frame'' in the NMR literature), which corresponds to 
a time-dependent interaction representation that follows the applied control. In this frame, the Hamiltonian is written as 
\begin{equation}
{\tilde H_0}(t)=U^{\dagger}_c(t)H_0 U_c(t),
\label{Hlogical} 
\end{equation}
where 
$U_c(t) = {\cal T} \exp [- i\int _{0}^t H_c (u) du]$ is the control propagator at time $t$, and the logical evolution operator becomes 
\begin{equation}
\tilde{U}(t)=U_c^{\dagger}(t)U(t)={\cal T} 
\exp \left[- i\int _{0}^t {\tilde H_0}(u) du \right] \:.
\label{frames}
\end{equation}

In this work, we shall focus on an isolated (closed) finite-dimensional system $S$, controlled through a sequence of equally spaced control pulses, $P_k$, applied at times $t_k$, $k \in \mathbb{N}$ $(t_0=0)$. The 
pulses average out the effects of unwanted interactions by repeatedly rotating the system and undoing its internal (drift) evolution.
In the limiting situation of arbitrarily strong and instantaneous
pulses -- the above-mentioned {bang-bang} setting \cite{Viola1998} -- 
the evolution during the pulses depends only on the control Hamiltonian,
whereas during the intervals $\Delta t = t_k - t_{k-1}$, the system evolves freely according to $H_0$. The propagator at $t_n=n \Delta t$, $n\in {\mathbb N}$, then reads 
\begin{eqnarray}
U(t_n)  
&=&P_n U(t_n,t_{n-1})P_{n-1} U(t_{n-1},t_{n-2}) 
\ldots P_1 U(t_1,0)P_0 \nonumber \\
&=& \underbrace{(P_n P_{n-1} \ldots P_1 P_0)}
\underbrace{(P_{n-1} \ldots P_1 P_0)^{\dagger} 
U(t_n,t_{n-1})
(P_{n-1} \ldots P_1 P_0)\ldots 
U(t_2,t_1)(P_1 P_0) P_0^{\dagger} U(t_1,0) P_0}\:. 
\label{U_physical} \\
&&\hspace{1. cm} U_c(t_n) \hspace{6.1 cm} \tilde{U}(t_n) \nonumber
\end{eqnarray}

The design of multi-pulse sequences is based on the desired form
of the effective propagator at a final evolution time $T>0$. To 
derive the time evolution operator, different methods have been 
employed, including {\em Fer's expansion}, which gives an exponential infinite-product expansion of $U(T)$ \cite{Fer1958,Klarsfeld1989,Blanes1998,Madhu2006}, 
and {\em average Hamiltonian theory} (AHT), which makes use of the
Magnus expansion to represent $U(T)$ in terms of a single exponential 
\cite{HaeberlenBook,ErnstBook}. Since the latter will be the main tool considered here, it is briefly described next.

We begin by writing the logical propagator at an arbitrary instant $t_n$
in terms of a single exponential. From Eq.~(\ref{U_physical}), 
\begin{eqnarray*}
\tilde{U}(t_n)
&=& 
\exp \left[-i ( P_{n-1} \ldots P_0)^{\dagger} H_0 (P_{n-1} \ldots P_0 ) \Delta t \right]
\ldots
\exp \left[-i (P_1 P_0)^{\dagger} H_0 (P_1P_0 ) \Delta t \right]
\exp \left[ -i P_0^{\dagger} H_0 P_0  \Delta t \right] 
\nonumber
\\
&=& \exp \left[-i H_n \Delta t \right] \ldots
\exp \left[-i H_2 \Delta t \right]
\exp \left[-i H_1 \Delta t \right], 
\nonumber \\
&=&\exp \left[ - i H_{eff}(t_n) t_n \right],
\label{Heff}
\end{eqnarray*}
where in the second line we have used the notation 
$H_n=( P_{n-1} \ldots P_0)^{\dagger} H_0 (P_{n-1} \ldots P_0 )$ 
for the transformed Hamiltonians during a given segment of evolution, and 
the Magnus expansion (or Baker-Campbell-Hausdorff expansion, since the Hamiltonian is piecewise constant in time) 
\cite{ErnstBook,HaeberlenBook,Klarsfeld1989PRA} to obtain 
the last equality. In the explicit expression of the effective
Hamiltonian $H_{eff} (t_n)=\sum_{k=0}^{\infty}{\bar
H}^{(k)} (t_n)$, each term ${\bar H}^{(k)} (t_n)$ is proportional 
to $(\Delta t)^k/t_n$ and involves $k$ time-ordered commutators of 
transformed Hamiltonians.

The convergence of the Magnus expansion depends strongly on the representation considered, and examples exist in the literature of its
failure at long times, see e.g. \cite{Maricq1982}. Explicit evaluations of the convergence radius have been obtained for specific systems, in particular for a two-level system 
\cite{Maricq1982,Feldman1984,Maricq1987,Salzman1987,Fernandez1990},
while general {\em sufficient} conditions for the absolute convergence of the expansion have been recently established, see e.g. \cite{Casas}
and references therein.
Interestingly, it has also been shown that by connecting the Magnus expansion with rooted trees, a recursive procedure to generate the expansion terms and a convergence proof become available \cite{Iserles1999,Iserles2002}.  For the current purposes, the condition 
$\kappa T_c<1$, where $\kappa=||H_0||_2=\max |{\rm eig} (H_0)|$
\cite{Maricq1982,ErnstBook,HaeberlenBook,Viola2005Random}, shall
be used as a guideline for the Magnus series convergence. 

So far, no special assumptions have been made in regard to 
the control field, which may, in principle, have either deterministic 
or non-deterministic features. While specific protocols within each setting will be described in Sec.~III, the essential difference between deterministic and randomized design is that in the latter case the {\em future} control path is not known, but rather, in the simplest case, effect a suitable random walk \cite{Viola2005Random}. In the particular case of a deterministic time-dependent perturbation which is {\em cyclic}, that is, when the control Hamiltonian and the control propagator are periodic with cycle time $T_c$, $H_c(t+n T_c)=H_c(t)$ and
$U_c(t+nT_c)=U_c(t)$, it follows from 
Eqs.~(\ref{Hlogical})-(\ref{frames}) that
the logical Hamiltonian is also periodic, and $H_{eff} (T_n)={\bar H}$ 
for any $T_n = n T_c$. At these instants, the system in the logical
frame appears to evolve under a {\em time-independent} 
average Hamiltonian ${\bar H}=\sum_{k=0}^{\infty}{\bar H}^{(k)}$,
the resulting propagator being 
$$\tilde{U}(nT_c)=\tilde{U}(T_c)^n=e^{-i {\bar H} n T_c}. $$
%
Accordingly, describing the system at any multiple integer of $T_c$ only requires the computation of the system's evolution after a single cycle. This constitutes the main result of AHT, and is also directly applicable to the physical frame: It follows from $U_c(0)=\mathbb{1}$ that
$U_c(nT_c)=\mathbb{1}$, which leads to the stroboscopic overlap of
physical and logical frames at $T_n$, $U(nT_c)=\tilde{U}(nT_c)$. 

For the deterministic DD sequences of relevance to this work, 
the first term of the average Hamiltonian ${\bar H}$, 
namely ${\bar H}^{(0)}$, may be cast in terms of a group-theoretic 
average \cite{Viola1}.  In this case, control pulses are successively drawn from a (projective) representation of a finite DD group 
${\cal G}=\{ g_j \}$, $j=0,\ldots, |{\cal G}|-1$, with $|{\cal G}|$ 
giving the order of the group.  The propagator after a control cycle, $t=T_c=|{\cal G}|\Delta t$, is written as
\begin{equation}
\tilde{U}(T_c) = \prod _{j=0}^{|{\cal G}|-1} U_{j+1}, 
\label{U_logical_PDD}
\end{equation}
where
\begin{eqnarray}
&&U_{j+1} = g_j^{\dagger} U(t_{j+1},t_j) g_j,
\hspace{0.5cm} H_{j+1}={g_j^{\dagger} H_0 g_j},
\hspace{0.5cm} 
P_{j+1}=g_{j+1} g_{j}^{\dagger}, \hspace{0.5cm} 
P_0=g_0  \: ,
\end{eqnarray}
The zeroth, first, and second order terms of the Magnus expansion are now 
respectively given by
\begin{eqnarray*}
&& {\bar H}^{(0)}  = \frac{\Delta t}{T_c} 
\sum_{k=1}^{|{\cal G}|} H_k ,
\label{Hprimeiro}
\\
&& {\bar H}^{(1)}  = -\frac{i(\Delta t)^2}{2T_c} 
\sum_{l=2}^{|{\cal G}|} \sum_{k=1}^{l-1} 
[H_l,H_k],  
\label{Hfirst} \\
&& {\bar H}^{(2)}  = -\frac{(\Delta t)^3}{6T_c}  \bigg\{
\sum_{m=3}^{|{\cal G}|} \sum_{l=2}^{m-1} \sum_{k=1}^{l-1}
\Big\{ [H_m,[H_l,H_k] + [H_m,H_l],H_k] \Big\}  
+ \frac{1}{2} \sum_{l=2}^{n} \sum_{k=1}^{l-1} 
\Big\{ [H_l,[H_l,H_k] + [H_l,H_k],H_k] \Big\}
\bigg\} . 
\label{Hsecond}
\end{eqnarray*}
In designing a DD scheme, one seeks an appropriate DD group ${\cal G}$ 
which may 'reshape' the target Hamiltonian as desired. The primary goal
is to tackle the dominant term ${\bar H}^{(0)}$, whose modification
may be sufficient in the ideal limit of $T_c\rightarrow 0$ or when dealing 
with very short evolution times. However, in realistic settings, and especially when long evolution times are involved, as in the current work, the role of higher order terms becomes critical, and strategies to reduce their effects are imperative. Among the various options, we shall discuss
symmetrization, concatenation, cyclic permutation, and randomization.

\subsection{Performance metric}

Our main control objective in this work will be to achieve a 'no-op' gate or, in NMR terminology, a time suspension -- that is, to freeze the system by completely refocusing the Hamiltonian evolution and making ${\tilde U}(T)$ as close as possible to the identity, $\mathbb{1}$, for a desired finite time $T$.  One way to quantify how successfully such objective is achieved relies on quantifying the input-output fidelity in the logical frame,
\begin{equation}
{\tilde F}_{\rho} (T)={\rm Tr}[{\tilde \rho}(T) \rho(0)] \;,
\label{fidelity}
\end{equation}
where $\rho(0)$ is an arbitrary initial state of the system and ${\tilde \rho}(T)={\tilde U}(T) \rho(0){\tilde U}^{\dagger}(T)$.  Arbitrary state preservation corresponds to the maximum value ${\tilde F}_{\rho} (T)=1$.
For a pure initial state $|\psi \rangle$, the above fidelity rewrites as
\begin{equation}
{\tilde F}_{|\psi \rangle} (T)=|\langle \psi |\tilde{U}(T)|\psi \rangle|^2.
\label{fidel}
\end{equation}

A disadvantage associated with ${\tilde F}_{|\psi \rangle} (T)$ is 
its intrinsic state dependence, a characteristic not suitable for a metric intended to assess dynamical protocol performance. In this sense, it would be more appropriate to invoke the pure state that leads to the worst-case pure state fidelity~\cite{Viola2005Random}. However, the drawback associated with this option is practical unfeasibility, since searching for the worst $|\psi \rangle$ is not operational, except for very small systems. Obtaining a control metric which is at same time state-independent and efficiently computable is possible by shifting attention from worst-case to typical input-state performance, as captured by so-called {\em entanglement fidelity}, $F_e$~\cite{Schumacher-Channels}.

Entanglement fidelity is defined with respect to an initial entangled state $|\psi^{RS}\rangle$ of the system $S$ and a reference system $R$
as $F_e(\rho^S,{\cal E}^S)={\rm Tr} \{|\psi^{RS}\rangle \langle \psi^{RS}| \rho^{RS'}\}$, where $\rho^{RS'}$ is the final state subjected to the evolution $\mathbb{1}^R\otimes {\cal E}^S$. By using the operator-sum representation, ${\cal E}^S=\sum_{\mu} A_{\mu}^S \rho^S A_{\mu}^{S \dagger}$, $F_e$ may be written in terms of quantities of the system only, $F_e(\rho^S,{\cal E}^S)=\sum_{\mu}|{\rm Tr} \{ \rho^S A_{\mu}^{S}\}|^2$. For a closed system, $A_{\mu}^{S}=U$, and the entanglement fidelity associated with a (any, in fact) maximally entangled purification $|\psi^{RS}\rangle$ -- thereby a maximally mixed state for $S$, $\rho^S=\mathbb{1}^S/d$ -- assumes the simple form $F_e(T)=|{\rm Tr}[ U(T)]/d|^2$ \cite{Fortunato-DFS}, where $d$ is the dimension of the system state space. Protocol performance may then be evaluated solely in terms of the system propagator. It is worth noting that a linear relationship exists between $F_e$ and the {\em average fidelity} ${\bar F}$ over all possible initial pure states, ${\bar F} = (d F_e +1)/(d+1)$, as formally established in Refs.~\cite{Horodecki1999,Nielsen2002} 

By its own nature, randomized DD methods involve various control realizations, each leading to a different value of fidelity.
Thus, control performance in this case is estimated in terms of an 
appropriate statistical average over individual results. In the logical frame, denoting by ${\mathbb E}$ the ensemble expectation over all control realizations, the expected entanglement fidelity is given by
\begin{equation}
{\mathbb E}   \{ {\tilde F}_e(T)\}
={\mathbb E} \{ |{\rm Tr}[ {\tilde U}(T)]/d|^2 \}\:.
\end{equation}
Complete refocusing then translates into achieving 
${\mathbb E}\{{\tilde F}_e(T)\} \rightarrow 1$. In the numerical Monte Carlo studies performed in what follows, ensemble expectation is replaced by the more viable statistical average over a sufficiently large sample of control realizations, which we designate 
by $\langle \langle {\tilde F}_e (T)\rangle \rangle$.

\subsection{Logical vs physical frame}

The logical representation is a convenient theoretical tool
used to facilitate the design of pulse sequences. However,
experiments are performed in the physical frame, so this is where 
our specific control objective need to be achieved. When dealing with periodic sequences, these differences are disregarded, since measurements are usually performed at the end of a cycle, where the two frames coincide. However, if one decides to observe the system in between cycles
or deals with {\em acyclic} pulse sequences (as it is inevitably the case in randomized DD), correcting pulses $P_c(t)$ may be required to guarantee
the final desired effect in the physical frame. It becomes then necessary to keep track of the applied pulses, because $P_c(t_n)$ at an arbitrary 
$t_{n}$ is determined by the control propagator $U_c(t_n)$ as
\begin{equation}
P_c(t_n)=U_c(t_n)^{\dagger}=(P_n P_{n-1} \ldots P_2 P_1 P_0)^{\dagger}.
\end{equation}

Consider, for example, the case of quantum information storage. 
Restricting ourselves to achieving $\tilde{U}(T)\rightarrow \mathbb{1}$ 
is equivalent, from the physical frame perspective, to
assuring that the system evolution is dictated only by the control, 
$U(T)\rightarrow U_c(T)$.  This is clearly reflected in Eq.~(\ref{fidelity}), which, by using Eq.~(\ref{frames}), may be rewritten as
\begin{equation}
{\tilde F}_{\rho} (T)={\rm Tr}[{\tilde \rho}(T) \rho(0)]
={\rm Tr}[ \rho(T) \rho_c(T)],
\end{equation}
where $\rho(T)=U(T) \rho(0) U^{\dagger}(T)$ and 
$\rho_c(T)=U_c(T) \rho(0) U_c^{\dagger}(T)$. 
Thus, to freeze the system in the physical frame also, conditional to a given control history, we need a correcting pulse that un-does $U_c(T)$, so that upon correction $F_{\rho} (T)={\rm Tr}[ \rho(T) \rho(0)] \rightarrow 1$, as desired.

Notice that in quantum information storage, frame correction and 
signal acquisition are performed only once, at the final time $T$. 
However, when data need to be constantly acquired, such as in standard line-narrowing NMR spectroscopy experiments, frequently applying 
frame-correcting pulses may be experimentally demanding, besides introducing additional errors. In such cases, it may be worth designing control schemes which need not be cyclic, but already incorporate
appropriate ``observation windows'' -- an example is given in Sec.~III.C.

\section{Dynamical Decoupling Design}

In this section, we outline several deterministic and randomized
DD schemes and discuss lower bounds for their attainable fidelity. Better performance depends on the protocol capabilities to increase averaging accuracy in the effective Hamiltonian and to slow down the accumulation of residual averaging errors. Symmetrization, concatenation, cyclic permutations, and randomization are the key design principles exploited to generate efficient schemes.

\subsection{Deterministic protocols}

We shall assume that the first group element for deterministic protocols is always $g_0=\mathbb{1}$, or equivalently, that the first pulse
occurs only after an initial time delay $\Delta t$.

(i) The simplest deterministic protocol is a cyclic scheme based on a {\em fixed}, {\em pre-determined} control path of a specific representation of ${\cal G}$, leading to first-order decoupling, ${\bar H}^{(0)}=0$. 
Any such scheme is referred to as {\em periodic DD} ({\tt PDD}). 
Following Eq.~(\ref{U_logical_PDD}), the logical propagator for {\tt PDD} at $T_c$ is given by
\[
\tilde{U}_{{\tt PDD}}(T_c) = \left( g^{\dagger}_{ |{\cal G}| -1} 
U (  |{\cal G}|\Delta t,(|{\cal G}|-1)\Delta t  ) g_{|{\cal G}|-1} \right) 
\ldots
\left( g^{\dagger}_{2}U (3\Delta t,2\Delta t) g_{2} \right)
\left(g^{\dagger}_{1} U (2\Delta t,\Delta t) g_{1}\right) 
\left( g^{\dagger}_{0}U (\Delta t,0)g_{0} \right), 
\]
which we compactly write as
\[
\tilde{U}_{{\tt PDD}}(T_c)=[U_{|{\cal G}|} \ldots U_3 U_2 U_1].
\]

From now on, $T_c$ will always refer to the cycle time of the {\tt PDD} sequence. Our goal, however, is to push beyond {\tt PDD}, by designing
deterministic protocols able to eliminate and/or reduce higher-order terms in the average Hamiltonian. Three main strategies are considered:

(ii) In analogy with the well-known Carr-Purcell sequence of NMR 
\cite{CP-Echo}, we may time-symmetrize the {\tt PDD} control path.
This leads to what we call {\em symmetric deterministic DD} ({\tt SDD}).  The cycle becomes twice as long, $T_c^{\tt SDD} = 2 T_c$, but
all odd order terms in ${\bar H_0}$ are also canceled~\cite{ErnstBook,HaeberlenBook}. In compact notation, the propagator becomes
\begin{eqnarray}
&{\tilde U}_{\rm {\tt SDD}} (2 T_c) = &
\underbrace{[U_1 U_2 U_3 \ldots U_{|{\cal G}|}]} \hspace{0.3 cm} 
[U_{|{\cal G}|} \ldots U_3 U_2 U_1].
\nonumber \\
&& \hspace{0.9 cm} [{\rm sym}] 
\nonumber
\end{eqnarray}

(iii) In {\em concatenated DD} ({\tt CDD}), the basic {\tt PDD} sequence
works as a ``seed" which is being recursively embedded within itself, as  formalized in~\cite{Khodjasteh2004,Khodjasteh2005}. At level of concatenation $(\ell +1)$, the pulse sequence in the physical frame
is determined by 
$C_{\ell+1}= C_{\ell}  P_1 C_{\ell} P_2  \ldots  C_{\ell} P_{|{\cal G}|} $, 
where $C_0$ denotes the interval of free evolution and $C_1$ is the generating inner {\tt PDD} sequence. Level $(\ell +1)$ is then reached at time $T=|{\cal G}|^{\ell} T_c$. In terms of group elements, since $g_0=\mathbb{1}$, we may write 
\[
{\tilde U}_{\rm {\tt CDD}_{({\ell+1})}} (|{\cal G}|^{\ell} T_c) =
\left(g_{|{\cal G}|-1}^{\dagger}{\tilde U}_{\rm {\tt CDD}_\ell}
g_{|{\cal G}|-1} \right)
\hspace{0.2 cm}
\ldots  \hspace{0.2 cm}
\left(g_2^{\dagger}{\tilde U}_{\rm {\tt CDD}_\ell} g_2 
\right)\hspace{0.2 cm}
\left( g_1^{\dagger} {\tilde U}_{\rm {\tt CDD}_\ell} g_1 
\right) \hspace{0.2 cm}
\Big( {\tilde U}_{\rm {\tt CDD}_{\ell}} \Big) .
\]
Note that at $\ell =2$, the above concatenated sequence is also symmetric, but, interestingly, it may outperform {\tt SDD} even before this level of
concatenation is actually completed, as analytically justified for the system considered here in Sec.~VI.A.3. This reflects {\tt CDD} efficiency in reducing the effects of higher order terms in the effective Hamiltonian.

Notice that if data is acquired before the completion of a given concatenation level, correcting pulses may be required to compensate for
frame mismatch. Besides, {\tt CDD} design is not cyclic. 
A periodic (or supercycle) version may be obtained by truncating the
scheme at a certain level $\ell$, and then periodically repeating it
at every $T=n|{\cal G}|^{\ell-1} T_c$ -- this protocol is denoted {\tt PCDD}$_{\ell}$.

(iv) Yet another alternative is inspired by the Malcolm Levitt's (MLEV) 
broadband decoupling sequence used in high-resolution NMR
\cite{Levitt1981,Levitt1982ab,Shaka1987}, which will be referred to as 
{\em symmetric cyclic permutation based DD} ({\tt SCPD}).
This pulse sequence combines symmetrization and cyclic 
permutations of the group elements in the following way. 
At what we call first level, m=1,  {\tt SCPD} and {\tt SDD} 
coincide. The cyclic permutations  initiate 
at level 2, being restricted to the {\tt PDD} part of the sequence as

\begin{eqnarray}
&{\tilde U}_{\rm {\tt SCPD}_2} (2 |{\cal G}| T_c) = &
\underbrace{[{\rm sym}]  [U_1 U_{|{\cal G}|} \ldots U_3 U_2]} \hspace{0.3 cm}
\ldots \hspace{0.3 cm}
\underbrace{[{\rm sym}]  [U_{|{\cal G}|-1} \ldots U_3 U_2 U_1 U_{|{\cal G}|}}] \hspace{0.3 cm}
\underbrace{[{\rm sym}]  [U_{|{\cal G}|} \ldots U_3 U_2 U_1 ]} \:,
 \nonumber \\
&&  \hspace{1.8 cm} A_{|{\cal G}|} 
\hspace{4.2 cm}  A_2 
\hspace{4.1 cm} A_1 
\nonumber
\end{eqnarray}
From the third level on, the sequence for m+1 is based on permutations of the entire sequence obtained at m, being concluded at 
$T=2 |{\cal G}|^{{\rm m}} T_c$. Following this rule, at m=3 we have
\begin{eqnarray}
&{\tilde U}_{\rm {\tt SCPD}_3} (2 |{\cal G}|^2 T_c) = &
\underbrace{  [A_1 A_{|{\cal G}|} \ldots A_3 A_2]} \hspace{0.3 cm}
\ldots \hspace{0.3 cm}
\underbrace{  [A_{|{\cal G}|-1} \ldots A_3 A_2 A_1 A_{|{\cal G}|}}] \hspace{0.3 cm}
\underbrace{ [A_{|{\cal G}|} \ldots A_3 A_2 A_1 ]} \:.
 \nonumber \\
&&  \hspace{1.0 cm} B_{|{\cal G}|} 
\hspace{3.8 cm}  B_2 
\hspace{3.6 cm} B_1 
\nonumber
\end{eqnarray}
Similarly to {\tt PCDD}$_{\ell}$, {\tt PSCPD}$_{\rm m}$ corresponds to 
a {\tt SCPD} sequence truncated at level m and periodically repeated at 
every $T=2n |{\cal G}|^{{\rm m}-1} T_c$.

A main disadvantage of periodically repeated sequences is that residual errors due to the higher-order terms in ${\bar H}$ accumulate coherently. However, this build-up slows down if the path to traverse ${\cal G}$ is constantly being changed, as indeed happens in both {\tt CDD} and {\tt SCPD}. This strategy is pushed to its limits by the use of randomization, as described next.

\subsection{Randomized protocols}

(i) The most straightforward randomized DD protocol is obtained by picking elements uniformly at random over ${\cal G}$ (notice that the relevant Haar measure is simply given by $1/|{\cal G}|$ in our discrete setting), such that the control action at each $t_n=n\Delta t$ ($t_0=0$ included)
corresponds to $P^{(r)}=g_i g_{j}^{\dagger}$, where $i,j=0,\ldots, |{\cal G}|-1$. This leads to the so-called {\em na\"{\i}ve random decoupling} ({\tt NRD}) -- an intrinsically acyclic method, which therefore prevents the direct use of AHT. The logical propagator at $T=n\Delta t$ for each of the $|{\cal G}|^n$ possible realizations is
\[
{\tilde U}_{\rm {\tt NRD}} (n \Delta t) = 
[ U_{r_n} \ldots U_{r_2} U_{r_1}], \hspace{0.3 cm} {\rm where} \hspace{0.3 cm}
r_1, r_2, \ldots, r_n \in R 
\hspace{0.3 cm} {\rm and} \hspace{0.3 cm} R=\{1,2, \ldots, |{\cal G}| \} .
\]

Comparing the two basic deterministic and randomized
schemes, {\tt PDD} and {\tt NRD}, the first is expected to 
perform better at short times, because it leads to 
${\bar H}^{(0)}=0$, whereas no guarantee exists of achieving
$H_{eff}(n|{\cal G}|\Delta t) \propto \Delta t$ with {\tt NRD}.
On the other hand, at long evolution times, {\tt NRD} is expected to
outperform {\tt PDD}, since it accumulates residual averaging errors
more slowly. To ensure good performance at both short and long times
it is then natural to seek for ways to merge advantageous deterministic
and stochastic features in a single DD scheme. With this goal in mind, we now describe several high-level alternatives for randomized protocols,
which may be thought as involving different choices for an ''inner" and an ''outer" control code \cite{Santos2006}. The inner code establishes 
the pulse sequence to be employed in certain intervals of the total final time and aims at increasing the minimum power of $\Delta t$ in the effective Hamiltonian, thereby improving short-time performance.
The outer code determines the random pulses applied at the borders of 
such intervals, with the objective of slowing down error accumulation.

(ii) A natural option corresponds to combining a fixed {\tt PDD} sequence used in the interval $[n,(n+1)]|{\cal G}|\Delta t$
with random pulses $P^{(r)}$ at $T_n=n|{\cal G}|\Delta t$. The bordering pulses may or may not be drawn from the same group ${\cal G}$. 
In the first case, {\em embedded DD 1} (${\rm {\tt EMD_1}}$), the
logical propagator at $T=|{\cal G}|\Delta t$ for each of the 
$|{\cal G}|$ possible realizations is
\[
{\tilde U}_{\rm {\tt EMD}_1} (|{\cal G}|\Delta t) = 
g_{j}^{\dagger}
[ U_{|{\cal G}|} \ldots U_3U_2U_1]g_{j} .
\]
As an example of the second case, we mention the protocol
implemented in~\cite{Kern2005}, here called ${\rm {\tt EMD}_2}$. 
The inner sequence corresponds to a {\tt PDD} based on a certain group ${\cal G}$, while the bordering pulses are drawn uniformly at random from the irreducible Pauli group ${\cal G}^P=\otimes_i {\cal G}_i$, where
${\cal G}_i = \{\mathbb{1}_i, X_i, Y_i, Z_i \}$, $i=1, 2, \ldots, N $, $N$ is the total number of two-level systems $i$, and $X_i, Y_i, Z_i$
are the Pauli matrices associated with each $i$. At $T=|{\cal G}| \Delta t$, there are $4^N$ possible realizations and the propagator for each one is given by

\[
{\tilde U}_{\rm {\tt EMD}_2} (|{\cal G}|\Delta t) =  
g_{p}^{\dagger}[ U_{|{\cal G}|} \ldots U_3U_2U_1]g_{p},  \hspace{0.3 cm} g_p \in {\cal G}^P \;.
\]
Since the number of realizations in ${\rm {\tt EMD}_2}$ is usually much larger than in ${\rm {\tt EMD_1}}$, error accumulation in the former is slower. In practice, however, situations may be encountered 
where control capabilities restrict protocol design to a single group.

(iii) The use of the {\tt PDD} sequence as the inner code guarantees only that an effective Hamiltonian with norm of ${\cal O}(\Delta t)$ is obtained. To ensure higher powers in $\Delta t$, we may embed with random pulses higher-level deterministic protocols, such as {\tt SDD}, {\tt PCDD}$_{\ell}$, and {\tt PSCPD}$_{\rm m}$, which lead to schemes respectively denoted here by {\tt ESDD}, ${\tt EPCDD}_{\ell}$, and ${\tt EPSCPD}_m$.

(iv) Another disadvantage of having a {\tt PDD} sequence as the inner code
is the fact that its performance may vary significantly depending on 
the specific path chosen to traverse ${\cal G}$. In cases where searching for the best option is costly, such as when ${\cal G}$ is large, a better
alternative consists in randomly choosing at every $T_n=n|{\cal G}|\Delta t$ a control path to traverse the group, leading to so-called {\em random path DD} ({\tt RPD})~\cite{Viola2005Random,Santos2006,Viola2006Snow}.
This scheme becomes yet more promising if the random paths are symmetrized in the same manner as in {\tt SDD}, leading to {\em symmetric random path 
DD} ({\tt SRPD})~\cite{Kern2006,Santos2006,Viola2006Snow}. The
logical propagator at $T=2 |{\cal G}| \Delta t$ for
each of the $|{\cal G}|!$ realizations is then given by
\begin{eqnarray}
&&{\tilde U}_{\rm {\tt SRPD}} (2|{\cal G}|\Delta t) = 
[{\rm sym}] [U_{s_{|{\cal G}}|} \ldots U_{s_3} 
U_{s_2} U_{s_1}], \hspace{0.3cm}  s_1 \in R, \; s_2 \in R-\{s_1\}, \;
\ldots  ,
\; s_{|{\cal G}|} \in R - \{ s_1,s_2,s_3, \ldots, s_{|{\cal G}|-1} \} 
\nonumber 
\end{eqnarray}

Since randomized protocols are intrinsically acyclic, correcting 
pulses are usually necessary before acquiring data. To avoid them, 
schemes which, as mentioned, may already contain suitable observation windows may be designed. As an example, we mention a pseudo-{\tt RPD}:
in this case, path randomization is restricted by the condition of having $U_0$ at every interval 
$[n|{\cal G}| \Delta t, (n |{\cal G}|+1) \Delta t]$, 
which ensures that physical and logical frame then coincide.

\subsection{Performance lower bounds}

Analytical bounds on the expected fidelity decay offer insight on relative
strengths and weaknesses of the proposed DD schemes.  Here, we both review existing error bounds and extend them to some of the new protocols of interest.

In the limit of sufficiently short time, $||H_0 ||_2 T<1$,
following Refs.~\cite{Viola2005Random,Kern2005} and expanding Eq.~(\ref{fidel}) to second order in $T$, the evolution in the logical frame of the fidelity for periodic DD may be written as
\[
{\tilde F}_{|\psi \rangle }(T) = |\langle 
\psi |e^{-i {\bar H} T} |\psi
\rangle|^2
\approx 1 - \big(\langle \psi |{\bar H}^2 |\psi \rangle -
\langle \psi |{\bar H} |\psi \rangle^2 \big)T^2 
+{\cal O}(T^3)\:.
\]
An upper bound for the square of the residual interaction 
$(\Delta {\bar H})^2 = 
\langle \psi |{\bar H}^2 |\psi \rangle -
\langle \psi |{\bar H} |\psi \rangle^2  $
is given by the norm of ${\bar H}$ as $(\Delta {\bar H})^2 \leq ||{\bar H}||_2^2$. In addition, the norm of the average Hamiltonian may also be bounded by $||{\bar H}||_2 \leq \sum_{j=0}^{\infty} \kappa (\kappa T_c)^j$, which finally leads to
\[
{\tilde F}_{|\psi \rangle }(T) 
\geq 1 - \bigg( \sum_{j=0}^{\infty} \kappa (\kappa T_c)^j 
\bigg)^2 T^2\:.
\]
A major factor influencing the performance of a deterministic protocol is its ability to suppress dominant terms in ${\bar H}$. Assuming that
the convergence condition $\kappa T_c<1$ is satisfied, and recalling the linear relation between ${\bar F}$ and ${\tilde F}_e$, we infer the following properties:

\begin{itemize}

\item {\tt PDD} cancels ${\bar H}^{(0)}$, therefore
$||{\bar H}||_2 \leq \kappa^2 T_c/(1 - \kappa T_c)$.
The limit $||{\bar H}||_2 T<1$ implies $\kappa^2 T_c T<1 - \kappa T_c$, which then leads to
${\tilde F}_e(T) \geq 1 - {\cal O}(\kappa^4 (|{\cal G}| \Delta t)^2 T^2)$.

\item {\tt SDD} cancels ${\bar H}^{(0)}$ and ${\bar H}^{(1)}$,
thus $||{\bar H}||_2 \leq \kappa^3 T_c^2/(1 - \kappa T_c)$,
thereby
${\tilde F}_e(T) \geq 1 - {\cal O}(\kappa^6 (|{\cal G}| \Delta t)^4 T^2)$.

\end{itemize}

The derivation of lower bounds for the performance of {\tt CDD} [{\tt SCPD}] is not straightforward, depending on three elements: the level of
concatenation [permutation], the model system, and the decoupling group 
considered. This is better discussed in Sec.~VI, where the dominant terms of ${\bar H}$ are explicitly computed for some particular models.
Here, we simply mention that when compared to {\tt PDD} and {\tt SDD},
{\tt PCDD} ($\ell>1$) and {\tt PSCPD} (m $>1$) are usually more efficient in reducing higher-order terms in the average Hamiltonian.

Contrasted with periodic methods, where residual errors due to higher order terms in ${\bar H}$ {\em build up coherently} (hence quadratically in time), the fidelity for random protocols decays linearly in time. This may be justified as follows. 

\begin{itemize}

\item Each step of {\tt NRD} can accumulate an error
amplitude up to $\kappa \Delta t$, and during a time $T$ there are
$T/\Delta t$ such intervals. Due to randomization, {\em amplitudes
add up probabilistically}, which leads to 
${\mathbb E}\{{\tilde F}_e(T)\}\geq 1 - 
{\cal O} \big( \kappa^2 \Delta t T \big)$.  The formal
derivation of this bound in the limit of $\kappa^2 \Delta t T \ll 1$ is presented in \cite{Viola2005Random}.

\end{itemize}

The reasoning is similar for the other protocols, although now each
step corresponds to the interval
$q|{\cal G}| \Delta t$, where $q=1$ for {\tt EMD}$_1$, 
{\tt EMD}$_2$, and {\tt RPD}, and $q=2$ for {\tt SRPD}. The bound 
becomes ${\mathbb E}\{{\tilde F}_e(T)\} \geq 1 - {\cal O}
( || H_{eff}||_2^2 |{\cal G}| \Delta t T)$, the norm 
of the effective Hamiltonian being an important difference between protocols. 

\begin{itemize}

\item {\tt EMD}$_1$, {\tt EMD}$_2$, and {\tt RPD} lead to
${\mathbb E}\{{\tilde F}_e(T)\}\geq 1 - 
{\cal O} \big( \kappa^4  (|{\cal G}| \Delta t)^3 T \big)$. 

\item {\tt SRPD} gives
${\mathbb E}\{{\tilde F}_e(T)\}\geq 1 - 
{\cal O} \big( \kappa^6  (|{\cal G}| \Delta t)^5 T \big)$.
The same lower bound holds for {\tt ESDD}, {\tt EPCDD}$_{\ell>1}$, 
and {\tt EPSCPD}$_{\rm m}$, although for the last two protocols 
averaging may be significantly better.  

\end{itemize}

In general, based on the above estimates, we then expect randomized methods to outperform their deterministic counterparts at long times.
For $T>(\kappa^2 |{\cal G}|^2 \Delta t)^{-1}$,
we should eventually have
${\mathbb E}\{{\tilde F}_e^{\tt NRD} (T)\}> {\tilde F}_e^{\tt PDD} (T)$, 
while $T>(\kappa^2 |{\cal G}| \Delta t)^{-1}$ leads to
${\mathbb E}\{{\tilde F}_e^{\tt EMD/RPD} (T) \} > {\tilde F}_e^{\tt SDD} (T)$.  However, in order to quantitatively compare {\tt SRPD} with {\tt CDD}, {\tt EPCDD}, {\tt SCPD}, and {\tt EPSCPD}, we need to specify the model in more detail. Notice that {\tt NRD} is the only protocol showing no dependence on the group size, which makes it a method of choice in 
cases where $|{\cal G}|$ is very large.

\section{Model System and Control Requirements}


\subsection{Model system}

We consider a chain with $N$ strongly coupled spin-1/2 particles (qubits)
described by the Heisenberg model, that is, the internal drift Hamiltonian in the physical frame reads
\begin{equation}
H_0=H_Z + H_{\rm int} = \sum_{i=1}^{N} \frac{\omega_i \sigma_i^{(z)}}{2}  + \sum_{i<j}^N \sum_{a=x,y,z} 
J_{ij}^{(a)} \sigma_i^{(a)}\otimes \sigma_j^{(a)} \:,
\label{ham}
\end{equation}
where $\sigma^{(a)}=\sigma^{(x,y,z)}=X,Y,Z$ are the Pauli operators, 
$\omega_i$ is the Zeeman splitting (Larmor frequency)
of spin $i$ as determined by a static magnetic field in the $z$ direction,
and $J^{(a)}_{ij}$ is the coupling parameter between spins $i,j$ in the $a$ direction. Open boundary conditions are assumed.

To illustrate the benefits of randomization, we concentrate on the simple case of homogeneous nearest-neighbor (NN) couplings, for which very efficient DD schemes exist (see Sec.~VI). By assuming 
$J_{ij}^x=J_{ij}^y=J$ and $J_{ij}^z=\alpha J$, where $\alpha$ is the coupling anisotropy associated with the Ising contribution,
we thus have:
\begin{equation}
H_{NN}= \sum_{i=1}^{N} \frac{\omega_i Z_i}{2} + 
\sum_{i=1}^{N-1} J \left[ X_i X_{i+1} + Y_i Y_{i+1} +\alpha Z_i Z_{i+1} \right] \:,
\label{hamNN}
\end{equation}
This Hamiltonian is used to model quasi-one-dimensional magnetic 
compounds~\cite{Salunke2007} and Josephson-junction-arrays~\cite{Glazman1997,Giuliano2005}.  It is also a fairly good approximation for couplings which decay exponentially with the qubit distance -- as arising, for instance, in semiconductor quantum dot arrays \cite{Loss}, or which decay cubically -- as in dipolarly coupled solid or liquid-crystal NMR spin systems~\cite{HaeberlenBook,LevittBook,Baugh2006} and electrons floating on Helium~\cite{Platzman,Dykman00}. 

Whenever qualitatively different, we shall compare the results associated with the above $H_{NN}$ with those obtained from cubically decaying interactions as approximated by the following Hamiltonian
\begin{equation}
H_{cub}= \sum_{i=1}^{N} \frac{\omega_i Z_i}{2} + 
 \sum_{i<j}^N J \left[ \frac{X_i X_j + Y_i Y_j +\alpha Z_i Z_j} {(j-i)^3} \right] \:.
\label{hamCub}
\end{equation}  
Although neglected here, an additional dependence on the angle between the vector joining spin pairs and the external magnetic field is present in principle in the secular dipole-dipole coupling parameter of NMR spin systems~\cite{HaeberlenBook,LevittBook}.

\subsection{Control requirements}

In order to suppress the interactions in Hamiltonians (\ref{hamNN}) and (\ref{hamCub}), we shall assume the ability to apply sequences of {\em selective} pulses, that is, control pulses that affect only some intended (subset of) spins. This is to be contrasted with {\em non-selective} (or collective) pulse sequences, which affect all qubits uniformly. A well-known example of the latter is the so-called WAHUHA (or WHH-4) sequence developed by Waugh, Huber, and Haeberlen~\cite{Waugh1968} to suppress direct dipole-dipole couplings.
A quantitative analysis of randomized versions of this sequence, which may have implications for solid-state NMR QIP, is left for a separate investigation.


Besides selectivity, another important feature of control pulses is the rotation angle they effect. Let us assume that, as it is the case in typical spin resonance experiments, 
the system couples to an oscillating control field linearly polarized in the $x$ direction according to
$$H_c(t) = 2\Omega (t) \cos [\omega_f t + \varphi (t)] 
\sum_{i=1}^{N} \frac{X_i}{2} \;,$$
where the amplitude (power) $2\Omega $, the carrier frequency $\omega_f$ and phase $\varphi$, as well as 
the interval $\tau$ during which $H_c(t)$ is on, and
the separation $\Delta t$ between successive pulses are under experimental control. The field is rapidly switched on and off so that $\Omega(t)$ may be approximated by a piecewise constant.  In the {rotating frame} of the carrier, which rotates with frequency $\omega_f$, the effective total Hamiltonian is given by
$$H^R(t)=U^{R\; \dagger}(t) \bigg(H_0 + H_c(t) -\frac{\omega_f}{2} \sum_{i}^{N}Z_i\bigg) U^R(t),
\hspace{5mm}
U^R(t)=\exp \bigg(-i\omega_f t\sum_{i}^{N}\frac{Z_i}{2} \bigg). $$
The interaction part of the Hamiltonian is invariant under this transformation, but, upon invoking the rotating wave approximation~\cite{HaeberlenBook}, the linear terms and the control Hamiltonian become
\begin{eqnarray*}
H_{Z}^R=\sum_{i=1}^N (\omega_i-\omega_f)\frac{Z_i}{2} 
=\sum_{i=1}^N \Delta_i \frac{Z_i}{2},    
\hspace{5mm}
H_c^R(t)=\Omega (t) 
\sum_{i=1}^{N} \left[ \frac{X_i}{2} \cos \varphi (t) 
+\frac{Y_i}{2} \sin \varphi (t) \right].
\end{eqnarray*}
From the above equations, we see that a given spin $i$ is rotated when the control field is applied on resonance with its frequency, 
$\omega_f \approx \omega_i$ (that is, the detuning $\Delta_i\approx 0$).
The phase $\varphi (t)$ then determines the direction around which
the rotation is realized in the rotating frame, and, in the case of rectangular pulses, $\Omega \tau$ characterizes the rotation angle. For instance, a pulse with $\omega_f=\omega_2$, $\varphi (t)=0$, and $\Omega \tau=\pi$ flips spin 2 by 180$^{\circ}$ around the $x$-axis. 
Here, the systems described by Eqs.~(\ref{hamNN}) and (\ref{hamCub}) 
will be subjected to sequences of $\pi$-pulses, while for instance the above-mentioned WAHUHA sequence involves $\pi/2$-pulses. 

All the analyses developed in this paper are performed in the rotating frame.
The spins in the systems of interest are assumed to be addressable in
frequency or by some other means, thereby the possibility of using selective pulses.  Additionally, the differences 
$|\omega_j - \omega_i|$ are assumed not to be much larger than the 
qubit-qubit coupling strength $J$, so that pulse sequences involving rotations around more than a single axis are required. If indeed 
$|\omega_j - \omega_i|\gg J$, the secular approximation leads to a 
truncated Hamiltonian where only terms in the $z$ direction remain 
\cite{LevittBook}. In this situation, DD may be effected by only using rotations around a single axis perpendicular to $z$
~\cite{Hahn-Echo,CP-Echo}. 
In the case of nuclear spin-1/2 Hamiltonians, this means that 
we are not interested in {heteronuclear} systems, since the Larmor frequencies of two different nuclear isotopes are separated by several 
MHz, while the couplings are of the order of tens or hundreds of Hz. Instead, our analysis has direct implications for {\em homonuclear} 
systems, where the spins are differentiated by their chemical shifts 
$\delta_i$, and $|\delta_j - \delta_i|\gtrsim J$. Chemical shifts emerge from the presence of electrons, which generate different small magnetic fields at different sites and cause variations of the net magnetic field experienced by the nuclei; the spin frequencies $\omega_0$ of the isotopes are then shifted as $\omega_i=\omega_0 + \delta_i$.

In short, we shall focus our analysis on the effects of DD sequences with multiple axes of rotation and selective pulses applied to the following rotating-frame Hamiltonians:
\begin{eqnarray}
&&H^{R}_{NN} \approx H_{\rm int} \;,
\label{HR-bi}
\\
&& H^{R}_{Z+NN} \approx \sum_i \frac{\delta_i Z_i}{2} + H_{\rm int} \;,
\label{HR-Zbi}
\end{eqnarray}
where in both cases $H_{\rm int}$ is given by the bilinear-NN-interaction terms described in Eq.~(\ref{hamNN}), and the two cases differ by the explicit inclusion of linear (chemical-shift) contributions. A comparison between the results for $H^{R}_{NN}$ and those for the cubic-decay couplings of Eq.~(\ref{hamCub}), $H^{R}_{cub}$, will also be provided.

\section{Randomization over Inefficient Decoupling Groups}

We begin by assessing the advantages of randomization in the case
of an inefficient DD group, that is, a group whose size increases exponentially with the number of qubits. In an appropriate logical-rotating frame, the system we consider is described by 
${\tilde H}^R_{NN}$ (\ref{HR-bi}).  Since ${\cal G}_k= \{{\mathbb 1}_k, Z_k, X_k, Y_k\}$ leads to {\tt PDD} sequences capable of refocusing  
$\sigma^{(a)}_{k-1} \cdot \sigma^{(a)}_{k}$ and $\sigma^{(a)}_k \cdot \sigma^{(a)}_{k+1}$, it is straightforward to see that 
${\cal G}=\otimes_k {\cal G}_k$, with $k=2,4, \ldots, 2m$ and $m\in {\mathbb N}$, may be used to obtain {\tt PDD} sequences which decouple up to $N$ qubits, where $N=2m$ or $N=2m+1$ \cite{Stoll,Viola1}. 
When $N=4$ or $5$, for instance, a possible DD scheme may be visualized in terms of the following matrix,
\[ M= \left(
\begin{array}{cccccccccccccccc}
\mathbb{1}& Z & X & Y 
& Y & X & Z & \mathbb{1}
&\mathbb{1}& Z & X & Y 
& Y & X & Z & \mathbb{1} \\
\mathbb{1}& \mathbb{1} & \mathbb{1} & \mathbb{1} 
& Z & Z & Z & Z
& X & X & X & X 
& Y & Y & Y & Y 
\end{array}
\right),
\]
where each row corresponds to an even qubit and each column,
supplemented with the identity operators associated to the odd qubits,
leads to an element of the group, so that ${\cal G}=\{g_j\}$,
$j=0,\ldots, |{\cal G}|-1$, with $g_j=\mathbb{1}_1 \otimes
[M_{(1,j+1)}]_2 \otimes \mathbb{1}_3 \otimes [M_{(2,j+1)}]_4
\otimes\mathbb{1}_5$. The proposed DD group requires $4^m$ 
$\pi$-pulses to close a single {\tt PDD} cycle. Any path taken to traverse ${\cal G}$ leads to first-order decoupling, however notice that a sequence arranged as in $M$ has the property of avoiding simultaneous rotations~\cite{Stoll}. Contrary to that, a path as in 
\[ M'= \left(
\begin{array}{cccccccccccccccc}
\mathbb{1}& Z & X & Y 
& \mathbb{1} & Z & X & Y
&\mathbb{1}& Z & X & Y 
& \mathbb{1} & Z & X & Y \\
\mathbb{1}& \mathbb{1} & \mathbb{1} & \mathbb{1} 
& Z & Z & Z & Z
& X & X & X & X 
& Y & Y & Y & Y 
\end{array}
\right),
\]
for example, leads to simultaneous rotations at every $t=4 n \Delta t$, 
$n\in \mathbb{N}$. Among all {\tt PDD} sequences derived from the above group ${\cal G}$, a very small subset consists of sequences involving only
single-qubit rotations; as $|{\cal G}|$ increases, most paths have in fact a large number of control actions involving simultaneous rotations on several qubits at a time.

In the {\tt NRD} protocol, which is based on uniform randomization over ${\cal G}$, possible control operations range from the total absence of rotations (the identity operator) to collective rotations on $m$ qubits at once. In large systems, the fraction of pulses corresponding
to extreme cases is very small. Let $Q_r=3^r m!/[r!(m-r)!]$ denote the total number of random pulses leading to $r$ simultaneous rotations for a given number $m$ of even qubits, where $r=0, 1, \ldots, m$, and $\sum_{r=0}^{m} Q_r =4^m$.  On the one hand, the percentage of pulses associated with a single qubit rotation, $r=1$, and with the maximum number of rotations, $r=m$, 
decreases with the size of the system -- as $3m/4^m$
and $(3/4)^m $, respectively.  On the other hand, the degree of parallelism increases significantly with $|{\cal G}|$.
Given $m$, the largest $Q_r$ is obtained for $r$ in the interval $[(3m-1)/4,(3m+3)/4]$ when $m\neq 3+4n$, whereas for $m = 3+4n$, both values, $(3m-1)/4$ and $(3m+3)/4$, lead to sets of equal size. This means that, for large $m$, the largest set of random pulses involves rotations on roughly $75\%$ of the even qubits.

\begin{figure}[th]
\psfrag{x}{$\hspace*{-4mm} {JT_n}$}
\psfrag{y}{$\hspace*{-8mm}{\langle \langle 
{\tilde F}^R_e \rangle \rangle }$}
\psfrag{z}{$\hspace*{-1mm} {Jt_n}$}
\begin{center}
\normalsize
\includegraphics[width=3.8in]{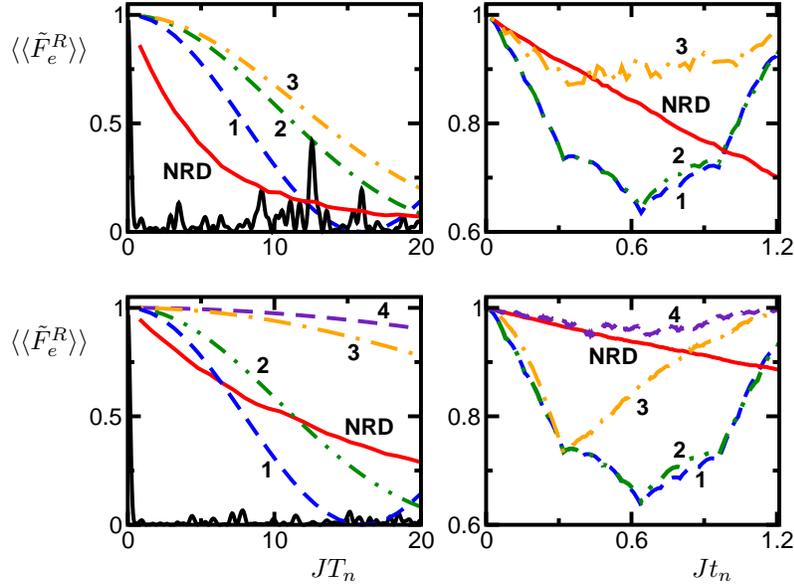}
\end{center}
\caption{ {\tt PDD} vs {\tt NRD} based on a nested pulse sequence for ${\tilde H}_{NN}^R$ (\ref{HR-bi}) with $\alpha =1$. Top panels: $N=6$, 
$|{\cal G}|=4^3$. Bottom panels: $N=8$, $|{\cal G}|=4^4$. Left panels: Ensemble-averaged entanglement fidelity at
$T_n=n|{\cal G}| \Delta t$; $\Delta t =0.8J^{-1}/|{\cal G}|$.
The numbers 1, 2, 3, and 4 stand for {\tt PDD}$_1$, {\tt PDD}$_2$, 
{\tt PDD}$_3$, and {\tt PDD}$_4$, respectively. Free evolution:
(black) oscillating solid line.  Right panels: 
$\langle \langle {\tilde F}_e^R \rangle \rangle$ at 
$t_n=n\Delta t<T_c$; $T_c=1.28J^{-1}$. 
Average over $10^2$ realizations.}
\label{fig:single}
\end{figure}

Whenever a high degree of parallelism is afforded, more efficient DD schemes exist where the total number of pulses needed to close a {\tt PDD} cycle is significantly reduced (see Sec.~VI).  However, the interest in the inefficient averaging schemes analyzed here lies on the possibility
to contrast the effects of single rotations versus simultaneous rotations, and to study DD under large control groups, while avoiding
computationally intractable system sizes. In Fig.~\ref{fig:single}, 
results on the decay of the ensemble-averaged entanglement fidelity in the rotating-logical frame, $\langle \langle {\tilde F}_e^R \rangle \rangle$, are shown. We consider $N=6$ ($N=8$) qubits in the top (bottom) panels, which leads to a relatively large control cycle: 64 pulses (256 pulses). In each column, both top and bottom panels have the same value
of $T_c$. We compare {\tt NRD} with different {\tt PDD} sequences: {\tt PDD}$_1$ --  based on the path given by $M$; {\tt PDD}$_2$ -- based
on the $M'$ path; and {\tt PDD}$_3$, which corresponds to a particular path selected at random and repeated at every $T_n=nT_c$. The beginning of the {\tt PDD}$_3$ sequence used on the bottom panel is equal to the {\tt PDD}$_3$ from the top panel. Another randomly selected path without this constraint is also considered in the case of $N=8$ and is referred to as {\tt PDD}$_4$.

When designing a {\tt PDD} sequence, it is natural to start with 
straightforward structures such as those given by $M$ or $M'$. However,
they are not necessarily the best options. In the left panels, fidelity
is computed at every $T_n=n |{\cal G}| \Delta t$. By contrasting top and bottom panels, we verify that {\em the performance of {\tt NRD} improves significantly as $|{\cal G}|$ increases}, while {\tt PDD}$_1$
and {\tt PDD}$_2$ remain essentially
unchanged. This explains the crossing between these two curves 
and the randomized protocol in the case of $|{\cal G}|=256$.
The strong enhancement resulting from parallelism becomes then evident and
suggests that better deterministic sequences ought to exist. In this sense, the selection of an efficient {\tt PDD} sequence is {\em a posteriori}  motivated by the study of a stochastic scheme.
In fact, {\tt PDD}$_3$ and {\tt PDD}$_4$ offer much better DD options.
In situations where different paths lead to such a broad range of performance and path optimization cannot be afforded, it is more
appropriate to use a protocol based on path randomization, such as {\tt RPD}. This scheme offers advantages also at long times, as it will be shown in Sec. VI.

In the right panels of Fig.~\ref{fig:single}, we also compare {\tt PDD} and {\tt NRD} during intra-cycle times, $t_n=n\Delta t$. This may be of interest in situations where constraints on the number of pulses or control intervals make it unfeasible to close a complete cycle, for
instance, when $T_c$ becomes prohibitively long. The decline in the {\tt PDD} performance for $t_n<T_c$ followed by its recovering as $t_n \rightarrow T_c$ reflects the fact that deterministic sequences are designed to perform well at the cycle completion, little being expected from them during intra-cycle times. Notice that up to half of the cycle,
{\tt NRD} is a protocol as good as, or even better than, the selected deterministic sequences.

We have then verified the beneficial contribution of parallelism in DD sequences, which is increasingly pronounced as the group size grows. However, to ''disentangle'' the two effects and isolate the impact of 
$|{\cal G}|$ in deterministic vs. randomized schemes, examining protocols
which have the same degree of parallelism is needed in principle -- e.g.,
those derived from combinatorics~\cite{Stoll}. The difficulty of such analysis, however, lies on the large system size required, which makes numerical simulations practically unfeasible.
 
As a further remark, we call attention to the cycle time used in the
figure: $T_c \sim J^{-1}$ is one order of magnitude larger than the values determined by the convergence criterion $\kappa T_c<1$. 
In the case of $N=8$, for instance, $\kappa \sim 20 J$. This confirms that the criterion is overly pessimistic in practice, and values of $T_c$ not necessarily complying with it may still lead to a substantial reduction of unwanted interactions in specific situations of interest.

\section{Randomization over Efficient Decoupling Groups}

We now focus on addressing the long-time behavior of the protocols described in Sec.~III. By long times we mean times where the analytical lower bounds are no longer reliable, $T\gtrsim (\kappa^2 \Delta t)^{-1}$.
Given the nearest-neighbor interactions under consideration, a very efficient DD group is now able to be identified, for which
{\tt PDD} always involves only $4$ selective multi-qubit pulses irrespective of system size.  Possible representations of 
the relevant control group for even $N$ are: 
\begin{eqnarray}
&&{\cal G}_{XY}=\{ \mathbb{1},\; X_1 X_3 \ldots
X_{N-1},\; X_1 Y_2 X_3 Y_4 \ldots X_{N-1} Y_{N},\; Y_2 Y_4 \ldots Y_{N}\} ,
 \nonumber \\
&&{\cal G}_{XZ}=\{ \mathbb{1},\; X_1 X_3 \ldots
X_{N-1},\; X_1 Z_2 X_3 Z_4 \ldots X_{N-1} Z_{N},\; Z_2 Z_4 \ldots Z_{N}\} ,
\nonumber \\
&&{\cal G}_{ZY}=\{ \mathbb{1},\; Z_1 Z_3 \ldots
Z_{N-1}, \; Z_1 Y_2 Z_3 Y_4 \ldots Z_{N-1} Y_{N}, \; Y_2 Y_4 \ldots Y_{N}\} ,
\label{repres}
\end{eqnarray}
where, in each case, the rotation axis for odd qubits is 
perpendicular to the rotation axis for even qubits. Notice that, if desired, the same averaging effects may be obtained with DD groups which affect only even or only odd qubits.  As an example, compare one of the pulse sequences derived from ${\cal G}_{XY}$: 
$P_1=P_3=X_1 Y_2 X_3 Y_4 \ldots X_{N-1} Y_{N}$
and $P_2=P_4=Y_2 Y_4 \ldots Y_{N}$, with a sequence
acting only on odd qubits: $P_1=P_3=Z_1 Z_3 \ldots Z_{N-1}$
and $P_2=P_4=Y_1 Y_3 \ldots Y_{N-1}$, which is derived from 
${\cal G}_{\rm{odd}}=\{ \mathbb{1},\; 
X_1 X_3 \ldots X_{N-1}, \; Y_1 Y_3 \ldots Y_{N-1}, \; 
Z_1 Z_3 \ldots Z_{N-1}, \}$. Both lead to the same transformed Hamiltonians in each segment of free evolution, and therefore
to the same results.

The small size of the DD group simplifies the derivation of the leading terms in the AHT, which, in turn, help anticipating the long-time behavior of the protocols.  In view of this, the strategy of this section is to first obtain and discuss the results for ${\bar H}^{(0)}$, ${\bar H}^{(1)}$, and ${\bar H}^{(2)}$ analytically, and then validate the analysis with numerical simulations.

\subsection{Analytical results}

For clarity, we show here the results obtained for a system described by ${\tilde H}_{NN}^R$, and leave the case where the linear chemical-shift Hamiltonian is retained, ${\tilde H}_{Z+NN}^R $, to the Appendix.

\subsubsection{Lowest-order average Hamiltonian}

At $T_n=nT_c =4n\Delta t$, first-order DD is achieved with 
any of the deterministic protocols, as discussed in Sec.~III,
\begin{equation} 
{\bar H}^{(0)}=\frac{H_1+H_2+H_3+H_4}{4}=0
\label{H0R}
\end{equation}
At these times, for all randomized protocols except {\tt NRD}, we also have, in the worst case, $H_{eff}(4n\Delta t) \propto {\cal O}(\Delta t)$.

\subsubsection{First-order contribution to the average Hamiltonian}

Using Eq. (\ref{H0R}), the first-order correction to the 
average Hamiltonian, ${\bar H}^{(1)} = -i(\Delta t)^2\{ [H_4,H_3] + [H_4,H_2] + [H_4,H_1] + [H_3,H_2] + [H_3,H_1]+ [H_2,H_1] \}/(2T_c)$, simplifies to
\begin{equation}
{\bar H}^{(1)} 
=-\frac{i(\Delta t)^2}{2T_c} \Big\{ [H_4,H_3]+[H_2,H_1] \Big\},
\label{Hone}
\end{equation}
whose result varies according to the group path selected. For each representation in Eq. (\ref{repres}), the $4!$ available paths lead to the following six different results, 

\begin{eqnarray}
&&{\bar H}^{(1)} = \pm J^2 \alpha \Delta t
\sum_{i=1}^{N-2} \left( Y_i X_{i+1} Z_{i+2}
+ Z_i X_{i+1} Y_{i+2} \right) ,
\label{H1Ax}
\end{eqnarray}
\begin{eqnarray}
&&{\bar H}^{(1)} = \pm J^2 \alpha \Delta t
\sum_{i=1}^{N-2} \left( X_i Y_{i+1} Z_{i+2}
+ Z_i Y_{i+1} X_{i+2} \right) ,
\label{H1Ay}
\\
&&{\bar H}^{(1)} = \pm J^2 \Delta t
\sum_{i=1}^{N-2} \left( X_i Z_{i+1} Y_{i+2}
+ Y_i Z_{i+1} X_{i+2} \right) . 
\label{H1Az}
\end{eqnarray}
Note that, in the three-body contributions appearing in 
${\bar H}^{(1)} $, the direction $a$ of the middle operator matches 
the direction of the interaction term 
$\sigma_i^{(a)}\otimes \sigma_{i+1}^{(a)}$ that most frequently (three times) changes sign within the interval $[0,4\Delta t]$. 
Therefore, for an anisotropic model with $\alpha>1$, paths that change the sign of the Ising term after every $\Delta t$, as in Eq.~(\ref{H1Az}), are preferable to those leading to Eqs.~(\ref{H1Ax}) and (\ref{H1Ay}), as intuitively expected.

In order to eliminate ${\bar H}^{(1)}$, we may employ symmetrized sequences such as {\tt SDD}, {\tt PCDD}$_{2}$, or {\tt PSCPD}$_{2}$. Specifically:

\begin{itemize}

\item The {\tt SDD} cycle consists of eight intervals of free evolution characterized by the transformed Hamiltonians in the following order,  $H_1, H_2,H_3,H_4,H_4,H_3,H_2,H_1$ -- or $(1234-4321)$ for short. 
The last four intervals correspond to a {\tt PDD} sequence where
$1\rightarrow 4$, $2\rightarrow 3$, $3\rightarrow 2$, 
and $4\rightarrow 1$, which inverts the sign of ${\bar H}^{(1)}$
in Eq.~(\ref{Hone}), leading to ${\bar H}^{(1)}(2nT_c)=0$.
Equivalently, for {\tt SRPD}, $H_{eff}(8n\Delta t) \propto {\cal O}
\left((\Delta t)^2\right)$.

\item {\tt PCDD}$_{2}$ is characterized by sixteen intervals of duration 
$\Delta t$, $(1234-2143-3412-4321)$, which is also symmetric, 
ensuring ${\bar H}^{(1)}(4nT_c)=0$. Interestingly, {\em half} of this
sequence also leads to ${\bar H}^{(1)}(2nT_c)=0$, since, according 
to Eq.~(\ref{Hone}), we can change the sign of ${\bar H}^{(1)}$ by 
simply switching the order in the pairs: $12 \rightarrow 21$
and $34 \rightarrow 43$.

\item {\tt PSCPD}$_{2}$ is given by the sequence 
$(1234-4321-4123-3214-3412-2143-2341-1432)$, so that after
every eight intervals $\Delta t$ we have ${\bar H}^{(1)}(2nT_c)=0$.

\end{itemize}

\subsubsection{Second-order contribution to the average Hamiltonian}

The three sequences given above do not cancel ${\bar H}^{(2)}$. 
In fact, even higher levels of concatenation (or permutations) are still incapable of eliminating the second order term in the AHT, due to the 
sequence pre-determined structure. The same $4$ ($8$) different paths employed in {\tt PCDD}$_{2}$ (or {\tt PSCPD}$_{2}$) are the only ones appearing also at $\ell>2$ (m$>2$), and whether alone or in rearranged combinations with each other, they cannot cancel ${\bar H}^{(2)}$.
This is to be contrasted with the sequence introduced in the end of this subsection, which incorporates a larger variety of group paths and does 
lead to ${\bar H}^{(2)}=0$. 
In order to better analyze the structure of ${\bar H}^{(2)}$, let us take advantage of Eq.~(\ref{H0R}) and write 
\begin{equation}
{\bar H}^{(2)} = -\frac{(\Delta t)^3}{6 T_c}
\Big\{ [(2H_1 + H_2),[H_1,H_2]] + [(2H_4+H_3),[H_4,H_3]] \Big\}.
\label{H2R}
\end{equation}
Because ${\bar H}^{(k)}=0$ for $k=0,1$, to obtain 
${\bar H}^{(2)}$ at $T_n$, we only need to sum ${\bar H}^{(2)}$ computed for each of the $n$ intervals $[0,4\Delta t]$. It is straightforward to verify that ${\bar H}^{(2)}$ obtained with a {\tt PDD} sequence is identical to the one computed with its corresponding {\tt SDD}, since
the result for $(1234)$ is equal to that for $(4321)$. Furthermore, the symmetry of Eq.~(\ref{H2R}) allows to simplify the computation of 
${\bar H}^{(2)}$ for {\tt PCDD}$_{2}$ and {\tt PSCPD}$_{2}$. 
For the first, we need to evaluate ${\bar H}^{(2)}$ only
for $(1234)$ and $(2143)$, whereas the latter requires the 
calculation of ${\bar H}^{(2)}$ for $(1234)$, $(4123)$, $(3412)$, and $(2341)$. Notice that, up to third order in the AHT, the same results 
are then obtained for this system with either {\tt PCDD}$_{2}$ or {\em half} of this sequence.  Even though ${\bar H}^{(2)}$ is the dominant term for {\tt SDD}, {\tt PCDD}$_{2}$, and {\tt PSCPD}$_{2}$,
the last two sequences lead to a significant improvement. This may be understood upon close inspection of a particular pulse sequence
based on ${\cal G}_{XY}$, characterized by the path
$\{ \mathbb{1},\; X_1 Y_2  \ldots X_{N-1} Y_{N} , \; X_1 X_3 \ldots X_{N-1} , \; Y_2 Y_4 \ldots Y_{N}  \}$ -- which leads to $P_1=P_3=X_1 Y_2 X_3 Y_4 \ldots X_{N-1} Y_{N}$ and $P_2=P_4=Y_2 Y_4 \ldots Y_{N}$. The following exact results are found: 

\begin{eqnarray}
&& {\tt SDD}: \hspace{0.2 cm} {\bar H}^{(2)} (2 T_c)=
-J^3  (\Delta t)^2 2 \alpha
\left\{ \frac{2}{3} \sum_{i=1}^{N-2}(X_{i} X_{i+2} + Y_i Y_{i+2} - 
2 Z_i Z_{i+2}) 
-\alpha \Big(Y_1 Y_2 + Y_{N-1} Y_N +2 \sum_{i=2}^{N-2} 
Y_i Y_{i+1} \Big)
\right.
\nonumber
\\
&& \left.
+\frac{1}{3} \sum_{i=1}^{N-3} \left[ 
X_i X_{i+2} \left( 2 Y_{i+1} Y_{i+3} - Z_{i+1} Z_{i+3} \right) + 
Y_i Y_{i+2} \left( 2 X_{i+1} X_{i+3} - Z_{i+1} Z_{i+3} \right) 
- Z_i Z_{i+2} \left(  X_{i+1} X_{i+3} + Y_{i+1} Y_{i+3} \right)
\right] \right.
\nonumber
\\
&& \left. +2\alpha \sum_{i=1}^{N-3} Z_i X_{i+1} X_{i+2} Z_{i+3}
\right\}
\nonumber
\end{eqnarray}
\begin{eqnarray}
&& {\tt PCDD}_2: \hspace{0.2 cm} 
\hspace{0.2 cm} {\bar H}^{(2)} (4 T_c)=
-J^3  (\Delta t)^2 2 \alpha
\left\{ \frac{2}{3} \sum_{i=1}^{N-2}(X_{i} X_{i+2} + Y_i Y_{i+2} - 
2 Z_i Z_{i+2}) 
\right.
\nonumber
\\
&& \left.
+\frac{1}{3} \sum_{i=1}^{N-3} \left[ 
X_i X_{i+2} \left( 2 Y_{i+1} Y_{i+3} - Z_{i+1} Z_{i+3} \right) + 
Y_i Y_{i+2} \left( 2 X_{i+1} X_{i+3} - Z_{i+1} Z_{i+3} \right) 
- Z_i Z_{i+2} \left(  X_{i+1} X_{i+3} + Y_{i+1} Y_{i+3} \right)
\right] 
\right\}
\nonumber
\\
&& {\tt PSCPD}_2: \hspace{0.2 cm} {\bar H}^{(2)} (8 T_c)=
+J^3  (\Delta t)^2  \alpha
\left\{ \frac{2}{3} \sum_{i=1}^{N-2}(Z_i Z_{i+2} + Y_i Y_{i+2} -2 X_{i} X_{i+2} ) 
\right.
\nonumber
\\
&& \left.
+\frac{1}{3} \sum_{i=1}^{N-3} \left[ 
 Z_i Z_{i+2} \left(  2 Y_{i+1} Y_{i+3} - X_{i+1} X_{i+3} \right)
+Y_i Y_{i+2} \left( 2 Z_{i+1} Z_{i+3}- X_{i+1} X_{i+3}  \right) 
-X_i X_{i+2} \left( Y_{i+1} Y_{i+3} + Z_{i+1} Z_{i+3} \right)
\right]
\right\} 
\nonumber
\end{eqnarray}
The results vary slightly for other control paths
(see the Appendix for a comparison between two possibilities), but the basic conclusion remains unchanged: the number of bilinear and four-body terms reduces when we switch from {\tt SDD} to {\tt PCDD}$_2$
or {\tt PSCPD}$_2$. In particular, notice that, contrary to {\tt SDD}, the bilinear terms in {\tt PCDD}$_2$ and {\tt PSCPD}$_2$ involve only
next-nearest-neighbor interactions.

In the case of ${\tilde H}_{Z+NN}^R$, where both linear and bilinear terms need to be taken into account, the outcomes for
${\bar H}^{(k)}$, $k=0,1,2$, become strongly dependent not only
upon the group path, but also on the representation chosen, as
demonstrated numerically in the next subsection and
analytically in the Appendix.

\subsubsection{Effect of group reducibility}

It is insightful to contrast the results of {\tt CDD} and {\tt SCPD} 
obtained here for the spin chain described by ${\tilde H}_{NN}^R$ with the case of a single qubit subject to a magnetic field of unknown direction, described by the Hamiltonian $H_0=\vec{B} \cdot \vec{\sigma}$. In both problems, the DD group consists of four elements, however for
${\tilde H}_{NN}^R$ the group action on the system's Hilbert space is {\em reducible}, whereas for the single qubit it is {\em irreducible}.  In the latter case, the irreducible decoupling group, 
${\cal G}=\{ \mathbb{1}, X, Y, Z\}$, is able to substantially decrease the power of $\Delta t$ in the average Hamiltonian for higher
levels of concatenation and permutation. The table below 
\cite{Pryadko_private} summarizes the order of ${\bar H}$ for the first four levels:
\begin{center}
\begin{tabular}{ccc}
& {\sf Isolated Single Qubit} & \\
Level & {\tt PCDD}$_{\ell}$ & {\tt PSCPD}$_{{\rm m}}$ \\ 
1 & ${\cal O}\big((\Delta t)\big)$ & ${\cal O}\big((\Delta t)^2\big)$\\
2 & ${\cal O}\big((\Delta t)^6\big)$ & ${\cal O}\big((\Delta t)^5\big)$\\
3 & ${\cal O}\big((\Delta t)^{23}\big)$  & ${\cal O}\big((\Delta t)^5\big)$ \\
4 & ${\cal O}\big((\Delta t)^{80}\big)$  & ${\cal O}\big((\Delta t)^5\big)$ 
\end{tabular}
\end{center}
As the level of concatenation increases, a superpolynomial convergence 
is verified, establishing {\tt CDD} as the best performer for this system.  For a single qubit coupled to an environment, the results depend fairly sensitively on the pure bath Hamiltonian~\cite{Khodjasteh2004,Khodjasteh2005}, which is renormalized by the control 
action~\cite{Wen03} and whose interplay with the system-bath coupling terms is responsible for determining the final convergence rate.
Still, provided that the environment dynamics is sufficiently slow, 
it has been verified that among the proposed protocols, {\tt CDD} 
remains the method of choice in the presence of generic single-qubit errors~\cite{Wen01,Wen02,Wen03}.

Having spelled out the advantages and limitations of {\tt CDD} and {\tt SCPD}, we now proceed to describe possible strategies to further improve protocol performance:

\begin{itemize}

\item One option, which is especially relevant for reducible DD groups, as in Eq.~(\ref{repres}), consists in truncating {\tt CDD} and {\tt SCPD} at the first level beyond which no further improvement is verified ($\ell=2$ and m=2 in the system under investigation), and then embedding the resulting periodic sequence with random pulses derived from an irreducible group, such as the Pauli group ${\cal G}^P=\otimes_i {\cal G}_i$. This way, the remaining terms in the effective Hamiltonian may still
be reduced.

\item  Another alternative is to take into account a larger number of group path realizations, and combine them into a supercycle sequence which, besides ${\bar H}^{(0)}$ and ${\bar H}^{(1)}$, also cancels 
${\bar H}^{(2)}$. This may be achieved, for instance, with the sequence $(1234-2143-2314-3241-3124-1342)$ -- see description below.
Once the appropriate sequence has been found, we may again exploit randomization and embed the supercycle with random pulses. 

\item Clearly, we may seek sequences which eliminate additional higher-order terms, although there may be in general some disadvantages associated with this: (i) the sequences may become much longer, and therefore harder to implement; (ii) searching for them may become very demanding, especially when dealing with complex systems and larger DD groups; (iii) in real settings, pulse errors need to be taken into account, which further significantly increases the complexity of the search problem.

\end{itemize}

\subsubsection{Supercycle sequence: 
${\bar H}^{(0)}={\bar H}^{(1)}={\bar H}^{(2)}=0$  }

In NMR, WAHUHA-based-supercycle sequences which are capable of eliminating dipolar interactions up to third order have long been devised \cite{HaeberlenBook}.  A simple approach consists in combining three WAHUHA sequences cyclically permuted~\cite{Mansfield1971}. In our case,
however, permutations of the basic path $(1234)$ are not sufficient, 
and more group path realizations are required. Indeed, 
${\bar H}^{(k)}=0$, $k=0,1,2$, may be achieved, for instance, with the sequence $(1234-2143-2314-3241-3124-1342)$.  Notice that each eight intervals of this scheme corresponds to a different 
half-{\tt PCDD}$_2$, which guarantees that ${\bar H}^{(k)}(2T_c)=0$ 
for $k=0,1$. Furthermore, by using Eq.~(\ref{H2R}) and adding the 
results for ${\bar H}^{(2)}$ obtained with each of the six {\tt PDD} sequences contained in the supercycle, we arrive at the desired result, ${\bar H}^{(2)}(6T_c)=0$. Thus, at every $T_n=6nT_c$, the first three terms in the average Hamiltonian are simultaneously canceled -- leading
to better averaging than for {\tt CDD} or {\tt SCPD} obtained in a cycle time even shorter than for {\tt PSCPD}$_2$.

In terms of pulses, this sequence, which we will refer to as 
${\tt H2}$ henceforth, translates into:
\[
U_c(6T_c)=P_A (P_C P_A P_C) P_B (P_C P_A P_C) P_C (P_B P_C P_B) P_A (P_B P_C P_B) P_B (P_A P_B P_A) P_C (P_A P_B P_A),
\]
where, for any path from Eq.~(\ref{repres}) which starts with the identity, that is, $\{\mathbb{1},g_1,g_2,g_3\}$, 
we have
$P_A=g_1=g_3 g_2^{\dagger}$, $P_B=g_2 g_1^{\dagger}=g_3$, and $P_C=g_3 g_1^{\dagger}=g_2$. 
Notice that the two axes of rotations involved in the basic first-order-DD sequences change every $8\Delta t$, and the direction appearing at every $4\Delta t$ alternates according to the following rule: it starts with $P_C$, is followed by $P_B$, and is finally $P_A$, being then 
repeated. This is to be contrasted with ${\tt PSCPD}_2$, where $P_C$ does not appear,
\[
(1234-4123-3412-2341)\Rightarrow U_c(4T_c)= 
\mathbb{1} (P_B P_A P_B)\mathbb{1} (P_A P_B P_A)\mathbb{1} (P_B P_A P_B)\mathbb{1} (P_A P_B P_A)\;,
\]
and with ${\tt PCDD}_2$, where $C_1$ is fixed, $P_C$ is the only rotation appearing in between two $C_1$'s,
and only $P_A$ or $P_B$ appear between $C_2$'s.
\[
C_2 \Rightarrow U_c(4T_c)= \mathbb{1} (C_1) P_C (C_1) \mathbb{1} (C_1) P_C (C_1),
\]
\[
C_3 \Rightarrow U_c(16T_c)= P_B (C_2) P_A (C_2) P_B (C_2) P_A (C_2).
\]

\subsection{Numerical results}

We validate the previous analytical analysis by studying a $N=8$ qubit system described by Eq.~(\ref{hamNN}), subject to selective DD pulses derived from Eq.~\ref{repres}. Whenever appreciably different, the results are also contrasted with those obtained for the cubically decaying Hamiltonian given by Eq.~(\ref{hamCub}).  Notice that in the latter case, DD sequences have been developed based on 
generalized Hadamard matrices~\cite{Stoll}, which may also be written in a group form as presented in~\cite{Santos2006}. For $N=8$ qubit, a possible representation is given by
\begin{eqnarray}
&{\cal G}_8=\{& \hspace{-0.1cm} \mathbb{1},Z_3 Z_4 Y_5 Y_6 X_7 X_8,Z_2 Y_3 X_4 Z_6 Y_7 X_8, 
Z_2 X_3 Y_4 Y_5 X_6 Z_7, 
\nonumber \\
&&
Y_2 Y_4 X_5 Z_6 X_7 Z_8, 
Y_2 Z_3 X_4 Z_5 X_6 Y_8,  
X_2 Y_3 Z_4 X_5 Z_7 Y_8,
X_2 X_3 Z_5 Y_6 Y_7 Z_8 \} \nonumber \:.
\end{eqnarray}

\subsubsection{Averaging of bilinear couplings: Isotropic system}

We first focus on the {\em bilinear} interaction terms alone, as 
in Eq.~(\ref{HR-bi}), with the main goal of comparing deterministic
and randomized protocols at long evolution times, where the
advantages of the latter are predicted to become important. 

\begin{figure}[th]
\psfrag{x}{$\hspace*{-2mm} { JT_n}$}
\psfrag{y}{$\hspace*{-8mm}\langle \langle 
{{\tilde F}^R_e}\rangle \rangle$}
\begin{center}
\includegraphics[width=4.5in]{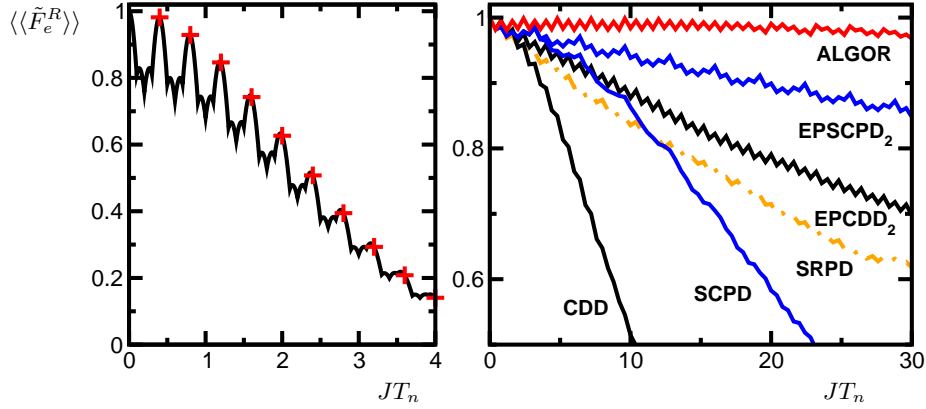}
\end{center}
\vskip - 2.5cm
\caption{Long time behavior of DD sequences based on
${\cal G}_{ZY}$ (\ref{repres}) and applied to ${\tilde H}_{NN}^R$ 
(\ref{HR-bi}) with $N=8$, $\alpha =1$, and $\Delta t=0.1 J^{-1}$.
Left panel: {\tt PDD} sequence. (Black) Curve: data acquired at intra-cycle times, $t_n=n\Delta t/5$; (red) crosses: data at $T_n=4n\Delta t$.
Right panel: Deterministic vs Randomized DD schemes. Data acquired 
at $T_n=4n\Delta t$. Average fidelity over $10^2$ control
realizations.}
\label{fig:NN8}
\end{figure}

As an initial illustration of the fast accumulation of errors occurring in periodic deterministic schemes, in the left panel of Fig.~\ref{fig:NN8} we assume a {\tt PDD} sequence and contrast the  
data acquired at intra-cycle times, $t_n=n\Delta t/5$, with data
obtained only at the completion of each cycle, $T_n=4n\Delta t$.
The intra-cycle curve oscillates in time. At short times, the peaks in 
performance coincide with the instants of cycle completion, but as time
evolves these two values become progressively detuned. This effect becomes more pronounced at longer times and for larger values of 
$\Delta t$, indicating that best performance is not necessarily achieved 
at $T_n$, and suggesting that repeating the same sequence after every 
cycle time may not be the best strategy.

We next proceed with a quantitative comparison between protocols.  While different ways for effecting such a comparison are conceivable, the most natural choice for contrasting cyclic and acyclic schemes is to fix the 
interval between consecutive pulses, implying that higher levels of concatenation and permutation may need longer times to be reached.  
Data is acquired after every $T_n=4n\Delta t$, which for some of the protocols provides information about the performance in between their defining inner sequences.  In the case of ${\tilde H}_{NN}^{R}$
and for times $T_n>30 T_c$, we find, in increasing order of 
performance: {\tt NRD}, {\tt PDD}, {\tt SDD}, {\tt EMD}$_1$, 
{\tt CDD}, {\tt EMD}$_2$, {\tt RPD}, {\tt SCPD}, {\tt SRPD}, 
{\tt EPCDD}$_2$, and {\tt EPSCPD}$_2$.  Since, with the exception of permutation-based protocols, these results have been already partially presented in \cite{Santos2006,Viola2006Snow}, we limit ourselves to showing in the right panel of Fig.~\ref{fig:NN8} the two best deterministic schemes and the three best randomized protocols, briefly commenting on the others in what follows.

(i) {\tt NRD} shows the poorest performance, consistent with the fact that the DD group is now very small and all protocols involve simultaneous rotations.

(ii) {\tt PDD} is unaffected by the representation or group path selected, whereas for ${\tilde H}_{cub}^R$, different choices lead to
a range of different results, which broadens as $|{\cal G}|$ increases.
Such a dependence also affects {\tt SDD} and randomized protocols
where the inner code is based on a fixed pulse sequence, such 
as {\tt EMD}.

(iii) {\tt EMD}$_2$ outperforms {\tt EMD}$_1$, which is not surprising given that the former involves an ensemble of $4^N$ random pulses, whereas the latter has only $4$. A comparison between {\tt RPD} and 
{\tt EMD}$_2$ is more subtle, due to the interplay between three factors: available repertoire of random pulses, chances for symmetrization being achieved at $T_n=8n\Delta t$, and sensitivity to path selection. For the
$NN$-isotropic system, {\tt RPD} shows the best performance, while for ${\tilde H}_{cub}^{R}$, specific path choices of the inner code lead to superior performance of {\tt EMD}$_2$ -- see Ref.~\cite{Santos2006}. 
In general situations where significant performance spread exist with respect to control path, even though superior {\tt EMD}$_2$ sequences
may exist, searching for them becomes demanding when $|{\cal G}|$ is large, which justifies the use of {\tt RPD} as a practical choice.

(iv) As seen in the right panel of Fig.~\ref{fig:NN8}, {\tt SRPD} surpasses first {\tt CDD} and then {\tt SCPD} at sufficiently long times. In contrast, for the system described by ${\tilde H}_{cub}^{R}$
with same $T_c$ value, {\tt CDD} is found to decay slower, being surpassed by {\tt SRPD} only at $T \geq 48 T_c$ (see Fig. 2 in \cite{Santos2006}), 
whereas {\tt SCPD} is outperformed by {\tt SRPD} already at $T\geq 4T_c$.
Still, the fact that such a simple sequence as {\tt SRPD} may outperform more elaborate deterministic methods such as {\tt CDD} and {\tt SCPD} vividly exemplifies the advantages of randomization.

\begin{figure}[thb]
\psfrag{x}{$ JT_n$}
\psfrag{y}{$\hspace*{-8mm}\langle 
\langle {{\tilde F}^R_e}\rangle \rangle$}
\begin{center}
\includegraphics[width=5.in]{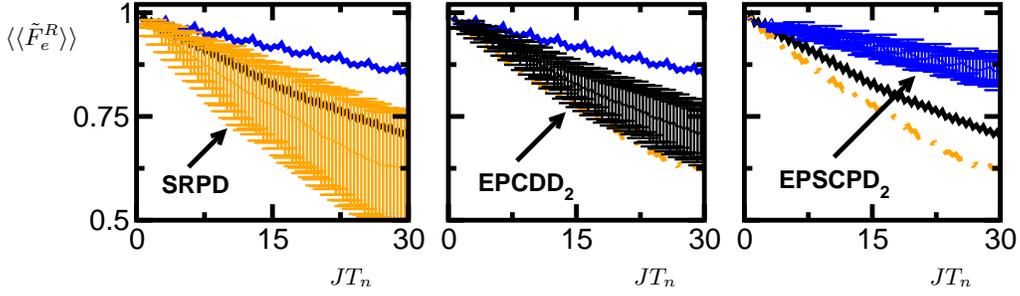}
\end{center}
\vskip -6cm 
\caption{Deterministic vs randomized DD schemes based on ${\cal G}_{ZY}$ (\ref{repres}) applied to ${\tilde H}_{NN}^R$ (\ref{HR-bi}) with $N=8$ and $\alpha =1$.  Data acquired at $T_n=4n\Delta t$, $\Delta t=0.1 J^{-1}$.  Average fidelity over $10^2$ control realizations.}
\label{fig:EMD}
\end{figure}

(v) The periodic sequences {\tt PCDD}$_{2}$ and {\tt PSCPD}$_{2}$  
embedded with pulses randomly picked from ${\cal G}^P$ perform better than {\tt SRPD}. In Fig.~\ref{fig:EMD}, we show the dispersions around the mean value for each of the three random schemes: as expected, they all broaden at longer times.  The best protocol, {\tt EPSCPD}$_{2}$, 
exhibits also the narrowest dispersion. 
Therefore, by combining randomization, symmetrization, 
and permutation, a DD scheme which is still relatively simple and yet efficient may be created.

\subsubsection{Sequence optimization}

Up to this point, the recipe we have been used to develop better DD protocols has consisted of deriving a first-order DD sequence, 
({\tt PDD}) from AHT, and then improving it by exploiting deterministic strategies and randomization.  We now address an alternative numerical approach to design high-level protocols. When creating algorithms to search for efficient protocols, the freedom in terms of types of controls (axis and angle of rotation), number of qubits affected at each step, and values of intervals between pulses is enormous, and taking all of these 
factors into account would make the analysis intractable.  Thus, in line with what we have done so far, we restrict to ideal selective pulses drawn from the sets in Eq. (\ref{repres}), and separated by a fixed interval $\Delta t$. 

The algorithm we propose may be described as follows. At every $T_{n}$, we search among the $|{\cal G}|!=24$ different group paths the one 
which leads to the largest value of
$\langle \langle {\tilde F}^R_e\rangle \rangle$ at $T_{n+1}$;
the best sequence from $T_{n}$ to $T_{n+1}$ is then stored, and
the same search procedure is iterated for the next intervals, so that 
the sequence is built up piece by piece. The resulting sequence is named
{\tt ALGOR} and is shown in the right panel of Fig.~\ref{fig:NN8}.
In a sense, this method shares some similarities with popularly employed genetic algorithms \cite{Mitchell96,Wu07}.  Here, the entire domain depends on the final time $T_n$, consisting of $(|{\cal G}|!)^n$ different individuals.  However, instead of randomly generating an initial population from this entire range of possible solutions, our
initialization is based only on the set of $|{\cal G}|!$ paths for the interval $[0,|{\cal G}| \Delta t]$. For each new interval 
[$T_{n}$, $T_{n+1}$] and with reference to the same population of 
$|{\cal G}|!$ paths, a new generation, bred from 
the best sequence for [$T_{n-1}$, $T_{n}$], is selected.
The fitness function corresponds to 
$\langle \langle {\tilde F}^R_e(T_{n+1})\rangle \rangle$: it 
strongly depends on time as well as on the previously selected ancestors. 
 
Below, we show the structure of the first $72$ intervals of free evolution for the optimized pulse sequence obtained with the parameters of Fig.~\ref{fig:NN8}:
\begin{eqnarray}
&&(1234-2143-2314-3241-3124-1342) \nonumber \\
&&(4312-4213-1423-4132-2431-3421) \nonumber \\
&&(4231-2413-4123-4321-3412-1432) 
\label{optSeq}
\end{eqnarray}
Observe that the first line corresponds to the scheme ${\tt H2}$ 
already discussed in Sec.~VI.A.5.
Each of the two additional lines in Eq.~(\ref{optSeq}) also
individually 
leads to the cancellation of ${\bar H}^{(0)}$, ${\bar H}^{(1)}$,
and ${\bar H}^{(2)}$.  The third-order decoupling is one reason 
for the significant improvement of this sequence when compared
with the others in Fig.~\ref{fig:NN8}. Another very important, and related, contributing factor is the uninterrupted variation of the control path at every $T_n=4n\Delta t$. Notice that up to $T=72 
\Delta t$, $18$ {\em different} control paths are used. This is to be contrasted with {\tt CDD} ({\tt SCPD}), where, for any level
of concatenation (permutation), only $4$ ($8$) different paths can be employed, variations being associated only with the order they are arranged.

\begin{figure}[h]
\psfrag{x}{$ JT_n$}
\psfrag{y}{$\hspace*{-8mm}\langle 
\langle {{\tilde F}^R_e}\rangle \rangle$}
\begin{center}
\includegraphics[width=3in]{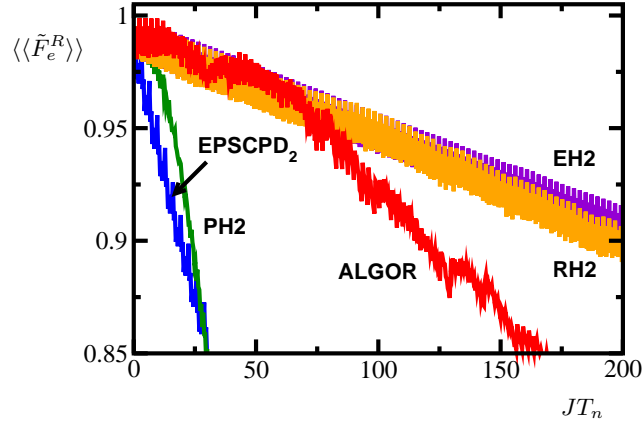}
\end{center}
\caption{DD schemes that guarantee
${\bar H}^{(0)}={\bar H}^{(1)}={\bar H}^{(2)}=0$ at $24 n \Delta t$ and {\tt EPSCPD}$_2$.  System described by ${\tilde H}_{NN}^R$
with $N=8$ and $\alpha =1$.  Data acquired at $T_n=4n\Delta t$, $\Delta t=0.1 J^{-1}$. Notation: {\tt ALGOR} -- sequence obtained via the numerical algorithm explained in the text; {\tt PH2} -- periodic sequence; {\tt EH2} -- embedded sequence with random pulses from 
${\cal G}^P$; {\tt RH2} -- random path.
Average fidelity over $10^2$ control realizations.}
\label{fig:optimal}
\end{figure}

Frequent path alteration is at the heart of methods employing path randomization, which makes it worth to further scrutinize the behavior of 
simpler sequences, whereby we use randomization on top of sequences achieving ${\bar H}^{(0)}(24n\Delta t)={\bar H}^{(1)}(24n\Delta t)={\bar H}^{(2)}(24n\Delta t)=0$.  Another motivation for this analysis is the fact that the algorithm proposed above clearly becomes unfeasible for large DD groups. In such cases, turning to simpler alternatives becomes a necessity.  Let us then select the first line in Eq.~(\ref{optSeq}) 
and create three new protocols: (a) a deterministic scheme where the $24$ free intervals are periodically repeated ({\tt PH2}); (b) a randomized scheme ({\tt RH2}), where the path for the interval 
$[24n\Delta t, 24n\Delta t + 4\Delta t]$ is picked at random 
and the subsequent interval 
$[24n\Delta t + 4\Delta t, 24n\Delta t + 24\Delta t]$
is rearranged so that at $24 (n+1) \Delta t$, the three
terms, ${\bar H}^{(0)}$, ${\bar H}^{(1)}$, and ${\bar H}^{(2)}$ cancel;
(c) another randomized scheme, ({\tt EH2}), where the 
first line is used as an inner code to be embedded with random pulses from ${\cal G}^P$. These protocols are compared in Fig.~\ref{fig:optimal} 
with {\tt ALGOR} and {\tt EPSCPD}$_2$.

Notice that the inner code of the two new randomized sequences is shorter
than that for {\tt EPSCPD}$_2$, yet they perform significantly better, {\tt EPSCPD}$_2$ being closer in performance to the deterministic scheme {\tt PH2}.  Interestingly, at very long times, {\tt RH2} and {\tt EH2}
outperform even the {\tt ALGOR} sequence. It is therefore clear that the algorithm used here cannot identify the best scheme for very long times,
the reason being the extreme sensitivity of pulse sequence performance to the final time. Take, for example, two instants of time $T_A$ and $T_B$, with $T_B>T_A$. The sequence that leads to the best result at $T_A$ is not necessarily the beginning of the one giving the best result at $T_B$. The algorithm employed looks for the best future pulses to be added to the paths that were already selected and which cannot be further altered.
The randomized sequences, on the other hand, have in storage realizations that may be worse than the {\tt ALGOR} scheme at $T_A$, but which will contribute to better realizations at $T_B$.

\subsubsection{Linear terms and anisotropy}

Attention so far has focused on averaging out the bilinear terms of an
isotropic system.  As a next step, we consider ${\tilde H}_{Z+NN}^R$,
by taking into account one-body terms and the effects of anisotropy. As a main feature, deterministic schemes (and by extension randomized 
schemes employing fixed inner codes) turn out to be strongly dependent upon the selected representation and control path. In such conditions, protocols based on path randomization become more advantageous for two main reasons: First, even though they need not lead to the best results, they ensure {\em robust behavior} against path variations; Second, it may simply be too demanding to find the best control path when dealing with large $|{\cal G}|$.  We shall compare two deterministic protocols, {\tt CDD} and {\tt SCPD}, with {\tt SRPD} in the presence of anisotropy and linear Zeeman terms characterized by $\delta_i$. The effective Hamiltonian for these three schemes is of ${\cal O}\big( (\Delta t)^2 \big)$, and exact analytical results for the second-order contribution ${\bar H}^{(2)}$ of the deterministic protocols are provided in the Appendix.

\begin{figure}[th]
\psfrag{x}{$\hspace*{-3mm} {JT_n}$}
\psfrag{y}{$\hspace*{-8mm}{\langle \langle {\tilde F}^R_e \rangle \rangle }$}
\begin{center}
\includegraphics[width=4.in]{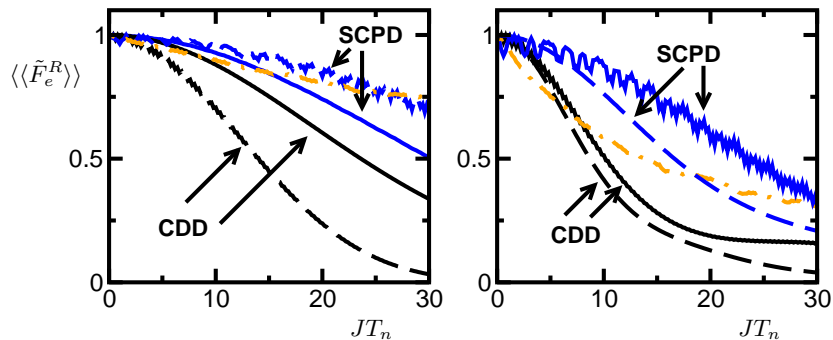}
\end{center}
\vskip - 1.0 cm
\caption{Deterministic vs randomized DD schemes based on  
${\cal G}_{XY}$ (\ref{repres}) applied to ${\tilde H}_{NN}^R$ and ${\tilde H}_{Z+NN}^R$, $N=8$. Data is acquired at every 
$T_n=4n\Delta t$, $\Delta t=0.05 J^{-1}$. For deterministic schemes: solid line -- Path 1; dashed line -- Path 2. {\tt SRPD}: dot-dashed line.
Left panel: ${\tilde H}_{NN}^R$; $\alpha =1$.
Qubits distinguished by chemical shift: $\delta_i=10 J$, $i$--odd; 
$\delta_i=0$, $i$--even.  Right panel: ${\tilde H}_{Z+NN}^R$
and $\alpha =5$. Average fidelity over $10^2$ control
realizations. Notice that the value of $\Delta t$ considered here
is half the one used in Figs.~\ref{fig:NN8}, \ref{fig:EMD}, reflecting the fact that, in the presence of $\delta_i $ and for $\alpha>1$, fidelity decays faster.}
\label{fig:aniso}
\end{figure}

Let us start by investigating the additional effects of the one-body terms in ${\tilde H}_{Z+NN}^R$. The selective pulses are now drawn from ${\cal G}_{XY}$ in Eq.~(\ref{repres}), since ${\cal G}_{XZ}$ and ${\cal G}_{ZY}$ do not cancel linear terms. Two paths are examined:
\begin{eqnarray}
&&{\rm Path}\; 1: \hspace{0.3 cm} \{ \mathbb{1},\; X_1 Y_2  \ldots 
X_{N-1} Y_{N} , \; X_1 X_3 \ldots X_{N-1} , \; Y_2 Y_4 \ldots Y_{N}  \},
\nonumber \\
&&{\rm Path}\; 2: \hspace{0.3 cm} \{ \mathbb{1},\; X_1 X_3 \ldots X_{N-1}  ,
\;  X_1 Y_2  \ldots 
X_{N-1} Y_{N}, \; Y_2 Y_4 \ldots Y_{N}\}.
\label{paths-aniso}
\end{eqnarray}
For $\delta_i > J \alpha$, based on  Eqs.~(\ref{H1Ay}), 
(\ref{H1Zee}), (\ref{path1}), (\ref{path2}), we expect Path $1$ to be the best choice for {\tt PDD}, {\tt SDD}, and {\tt CDD}, whereas Path $2$ is more suitable for {\tt SCPD}.  This is demonstrated numerically for {\tt CDD} and {\tt SCPD} in the left panel of Fig.~\ref{fig:aniso}, where an isotropic system, $\alpha=1$, is considered. Once again, the randomized scheme, {\tt SRPD}, surpasses the deterministic protocols at sufficiently long times. In the case where $\delta_i \lesssim J \alpha$, 
the competition between $\alpha $ and $\delta_i$ complicates the selection of the best control path, which encourages the use of randomized-path schemes.

In order to isolate the effects of the anisotropy, we discard the one-body terms and return to ${\tilde H}_{NN}^R$, but this time with $\alpha \neq 1$.  As indicated by Eqs.~(\ref{H1Ay}), (\ref{H1Az}), (\ref{path1}),
and (\ref{path2}), {\tt PDD} and {\tt SDD} are expected to perform
better for Path $2$, which is somehow intuitive, since this is the path that changes the sign of the Ising term more frequently. Predictions of this sort become less trivial when dealing with {\tt CDD} and {\tt SCPD}, since ${\bar H}^{(2)}$ for the interactions alone is very similar for both paths. Therefore, the identification of the best path for these
protocols requires either precise knowledge about the system and
tedious computations of higher-order terms in the average Hamiltonian, or a numerical search over an ensemble of realizations. Both options may be avoided if instead we employ a randomized protocols such as  {\tt SRPD}. 
In the right panel of  Fig.~\ref{fig:aniso}, we compare the paths from Eq.~(\ref{paths-aniso}) for an anisotropic system controlled via {\tt CDD}, {\tt SCPD}, and {\tt SRPD}. In stark contrast to {\tt PDD} and {\tt SDD}, {\tt CDD} and {\tt SCPD} perform better for Path $1$. Notice also that {\tt CDD} appears to be more robust than {\tt SCPD} against path variations; still, as before, they are both surpassed by {\tt SRPD} at long times. 

Overall, the following conclusions may be drawn: Various analytical and numerical strategies exist or may be devised to improve the performance of deterministic protocols.  However, DD may always benefit from randomization in terms of: pulse sequence simplification, robustness to path variations, and slower accumulation of residual averaging errors.

\section{Pulse Imperfections}

Throughout the previous analysis, we have assumed perfect control resources, implying, in particular, the ability to effect perfect instantaneous pulses.  In practice, attainable control operations are 
far from ideal, a variety of both systematic and random imperfections contributing to deteriorate protocol performance.  Systematic errors, 
in particular, may be especially harmful at long times, since their
effects tend to be cumulative.  Depending on implementation detail, different control non-idealities may be relevant ~\cite{HaeberlenBook}, including: finite-width effects; deviations from the intended rotation angles, which may in turn be common to all pulses or different for different sets of controls; phase errors, arising from the fact that 
the phases of different pulses are not necessarily in quadrature;
phase transients associated with control switching.
By way of illustration, we focus on analyzing how DD performance is affected by pulses of finite duration and flip-angle errors. 
The three protocols with effective Hamiltonian of 
${\cal O}\big( (\Delta t)^2 \big)$, {\tt CDD}, {\tt SCPD}, and 
{\tt SRPD}, are selected for such investigation, some discussion about {\tt PDD} being also presented. The case of a system described by 
${\tilde H}_{NN}^R$ is explicitly considered, with DD pulses being 
drawn from ${\cal G}_{ZY}$.

\subsection{Finite pulse widths}

In realistic control settings, the power $\Omega$ is not infinite nor is the pulse duration $\tau$ equal to zero. As a first approximation,
pulses may be assumed to have a rectangular profile (for shaped pulses see e.g. Refs.~\cite{SlichterBook,Vandersypen05,Sengupta05,Pasini07}),
and phase transients associated with the instants they are turned on and off \cite{HaeberlenBook} may be disregarded, so that the desired rotation angle is simply determined by the product $\beta=\Omega \tau$. 

\begin{figure}[th]
\psfrag{h}{$\hspace*{-4mm} {\beta/\pi}$}
\psfrag{z}{$\hspace*{-8mm}
{\langle \langle {\tilde F}^R_e \rangle \rangle }$}
\psfrag{x}{$\hspace*{-2mm} {JT_n}$}
\psfrag{y}{$\hspace*{-8mm}{\langle \langle {\tilde F}^R_e \rangle \rangle }$}
\begin{center}
\includegraphics[width=4.6in]{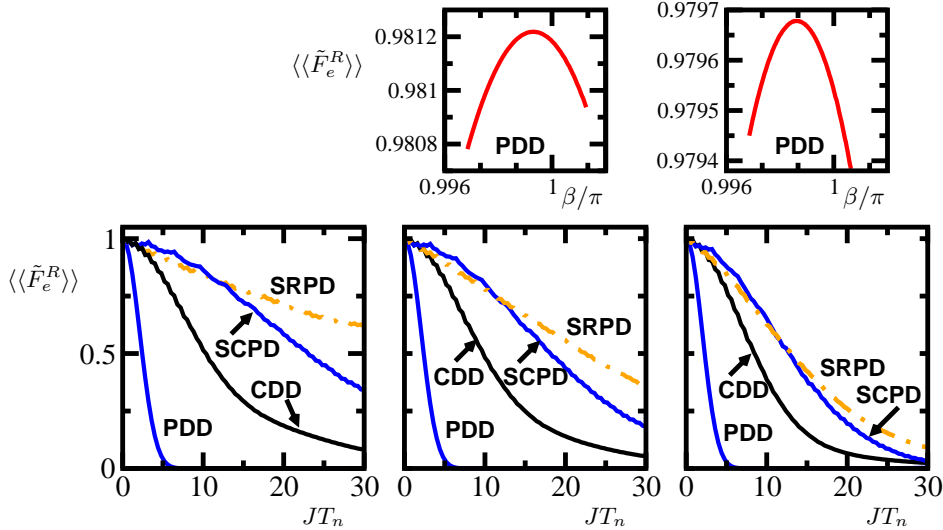}
\end{center}
\caption{Ideal vs finite-width pulses for deterministic and randomized DD schemes derived from ${\cal G}_{ZY}$ (\ref{repres}). System described by ${\tilde H}_{NN}^{R}$ with $N=8$, $\alpha =1$, and $\Delta t=0.1 J^{-1}$. 
Top panels: $\langle \langle {\tilde F}^R_e (T_c) \rangle \rangle $
vs $\beta/\pi$. Bottom panels: Decay in time for
$\langle \langle {\tilde F}^R_e \rangle \rangle $ for
$\beta =\pi$. Left panel: $\tau=0$. Middle panels: $\tau=0.005 J^{-1}$.
Right panels: $\tau=0.01 J^{-1}$. Average fidelity over $10^2$ control
realizations.}
\label{fig:finite}
\end{figure}

In the presence of finite pulses, first-order DD is no longer achieved. Instead, after the completion of the first {\tt PDD} cycle, we find
\begin{equation}
{\bar H}^{(0)}=-\sum_{i=1}^{N-1}
(Y_iX_{i+1}+X_iZ_{i+1})\frac{\tau (1 - \cos \beta)}{2\beta \Delta t}
\stackrel{\beta=\pi}{\rightarrow} -\sum_{i=1}^{N-1}
(Y_iX_{i+1}+X_iZ_{i+1})\frac{\tau}{\pi \Delta t},
\end{equation} 
which cancels only in the limit $\tau/\Delta t \rightarrow 0$.
This is to be contrasted with the WAHUHA sequence, where first-order DD may still be achieved by properly adjusting the rotation angle according to $\tau/\Delta t$ \cite{HaeberlenBook}. In our case, depending on such a ratio, small deviations from $\beta=\pi$ lead simply to hardly perceptible improvements on the results for ${\tilde F}^R_e(T_c)$, 
as shown in the top panels of Fig.~\ref{fig:finite}.  To justify this improvement, higher-order terms in the average Hamiltonian are needed, since it is most probably caused by the interplay between these terms and ${\bar H}^{(0)}$.

\vspace*{5mm}

Since, for the values of $\tau/\Delta t$ considered here, the improvement in fidelity obtained by varying $\beta$ is negligible, in the bottom panels of Fig.~\ref{fig:finite} we simply fix $\beta=\pi$ and
compare {\tt PDD}, {\tt CDD}, {\tt SCPD}, and {\tt SRPD}.
Similarly to Fig.~\ref{fig:NN8}, {\tt SRPD} outperforms the 
deterministic schemes at long times.  However, {\tt SRPD} deteriorates faster with finite pulses than the deterministic schemes. As a result, 
for very large errors, of the order of $\tau/\Delta t \geq 10\%$,
the gain achieved with randomization is offset by the errors and 
the performance of {\tt SRPD} becomes comparable to that of
{\tt SCPD}.

\subsection{Flip angle errors}

Flip angle errors may be caused by power misadjustment in the pulse generator, variations of the transmitter power output, or radio-frequency inhomogeneities~\cite{HaeberlenBook}.  Here, we focus on systematic flip angle errors which are common to all pulses.  This corresponds to a small over-rotation $\epsilon $ of the intended $\pi$-pulses, and is described by
\begin{equation}
\exp [-i\pi (1 + \epsilon) \sigma_i^{(a)}/2]=
-\mathbb{1}\sin (\epsilon \pi/2)
-i\sigma_i^{(a)} \cos (\epsilon \pi/2) \;.
\label{error}
\end{equation}

\begin{figure}[ht]
\psfrag{x}{$\hspace*{-4mm} {JT_n}$}
\psfrag{y}{$\hspace*{-8mm}{\langle \langle {\tilde F}^R_e \rangle \rangle }$}
\psfrag{z}{$\hspace*{-2mm} \vspace{20cm} {D(\epsilon)}$}
\begin{center}
\includegraphics[width=4.6in]{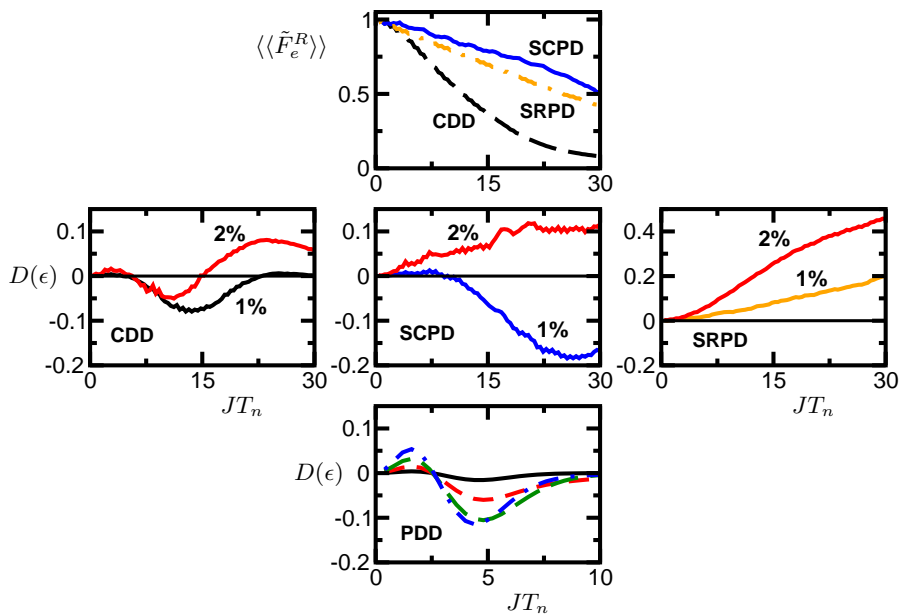}
\end{center}
\caption{Deterministic vs randomized DD derived from ${\cal G}_{ZY}$.
System described by ${\tilde H}_{NN}^{R}$, $N=8$, $\alpha =1$, and $\Delta t=0.1 J^{-1}$. Top panel: $\epsilon=0.01$.
Middle panels: 1\% and 2\% stand for $D(0.01)$ and $D(0.02)$,
respectively. Bottom panel: $D(\epsilon)$ for $\epsilon=0.01$ (black solid line) $\epsilon=0.02$ (red short-dashed line),
$\epsilon=0.03$ (green long-dashed line), and $\epsilon=0.04$ (blue dot-dashed line). Average fidelity over $10^2$ control realizations.}
\label{fig:faulty}
\end{figure}

In the top panel of Fig.~\ref{fig:faulty}, we consider an over-rotation of $1\%$ (which is relatively larger than what may be found for instance in typical resonance experiments~\cite{HaeberlenBook,Morton2005}), and compare {\tt SCPD}, {\tt CDD}, and {\tt SRPD}.  As in the case of ideal pulses, {\tt CDD} is outperformed by {\tt SRPD}, however the crossing between  {\tt SCPD} and {\tt SRPD} is no longer verified. This may be better understood by observing the middle panels, where we show the
difference $D(\epsilon)$ between the fidelity obtained
with ideal and with faulty DD pulses,  
$\langle \langle {\tilde F}_e^R \rangle \rangle_{\epsilon=0}$
and $\langle \langle {\tilde F}_e^R \rangle \rangle_{\epsilon}$, respectively.  Interestingly, errors may contribute {\em favorably}
to the performance of deterministic schemes, as indicated by the 
negative values of $D(\epsilon)$. When $\epsilon =0.01$, the improvement for {\tt CDD} is modest and occurs at intermediate times, while {\tt SCPD}
shows a significant increase in fidelity at long times.  Contrary to that,
flip-angle errors have always a detrimental impact on randomized schemes.
Therefore, the accumulation of high-order terms of the average Hamiltonian
in {\tt SCPD} is counterbalanced with the positive effects caused by errors $\epsilon \sim 0.01$, while the advantages of randomization in {\tt SRPD} cannot compensate for the sensitivity to pulse imperfections,
resulting in the worse performance of the latter.

As a further illustration of the effect of flip-angle errors in deterministic schemes, we show in the bottom panel the effects of $\epsilon$ on {\tt PDD}. At very short times, $\epsilon$ enhances the fidelity decay, while this situation is reversed at longer times.
Contrary to {\tt SCPD} and {\tt CDD}, where $\epsilon >0.01$ mostly worsens protocol performance,  a consistent improvement of {\tt PDD} at longer times is observed for errors up to $\epsilon \sim 0.03$.

In short, even though deterministic protocols appear to be more
protected against finite width and flip-angle errors than randomized
schemes, in the case of relatively small errors the advantages of randomization at long times are still dominant.  From this perspective, 
a promising next step may arise from combining randomized with bounded-strength Eulerian design \cite{Viola2003Euler}, which is explicitly 
intended to compensate unwanted evolution during pulses and offer enhanced fault-tolerance. 

%

\section{Conclusions}

\subsection{Summary}

We have developed a quantitative comparison between deterministic and randomized DD protocols in closed systems described by a time-independent Hamiltonian, confirming the advantages of randomization at long evolution times and the efficiency of control protocols which combine multiple 
decoupling strategies -- such as randomization, symmetrization, concatenation, and cyclic permutations.  We have also argued how the search for better deterministic sequences in a large set of possibilities may be shortcut by using randomization to develop simple, yet very efficient protocols.  While the main emphasis has been on removing bilinear interactions in a spin-$1/2$-particle-system with isotropic
$NN$ couplings, a number of results in the presence of anisotropic couplings and one-body terms have also been established.  Furthermore, 
a comparison between DD results for $NN$ and for long-range cubically decaying interactions has been included.  Two types of DD groups have been considered: an inefficient group whose size increases exponentially with the system size, and may be easily extended to systems with long range couplings; and a very efficient group, which leads to only four
simultaneous pulses and is specifically designed to systems with $NN$ interactions.

In the case of inefficient averaging, we have shown that different paths to traverse the DD group lead to a broad range of results, where
{\tt PDD} sequences involving collective rotations tend to perform better than those consisting mainly of single rotations. 
For large groups, the selection of the best deterministic protocols
becomes very demanding, which favors protocols that average over
various possibilities, such as {\tt NRD}. One step further consists in applying {\tt RPD}, which already pre-selects the most efficient pulse sequences to be included in the average.  Additionally, we have showed that in situations where the DD group is so large that a single cycle can be hardly completed, the performance of {\tt NRD} is similar to the best {\tt PDD} performance.
 
The small number of pulses involved in efficient DD schemes has  
allowed for a thorough analytical study. 
This has offered insight into understanding why different paths and 
group representations do not affect DD performance in the case of isotropic $NN$ couplings, and also to partially predict the best control choices in the presence of anisotropy and one-body terms -- which have 
been next numerically validated.  Most importantly, the analytical results
have shed light on the reasons for the limited performance of concatenated protocols (and protocols based on cyclic permutations) in the class of systems under consideration, and paved the way to the development of a better control sequence able to decouple interaction up to third order, at least. The key idea has been to access more path realizations than those available to {\tt CDD} and {\tt SCPD}, and yet rearrange them such that the structure of half-${\tt PCDD}_2$ was kept.

Numerical simulations have served a twofold purpose: to confirm
and extend the analytical predictions; and to identify the best
randomization strategy. While randomization is unquestionably advantageous at long times, whether it is better to embed a deterministic sequence with random pulses or to apply path randomization strongly depends on the system at hand. If the inner code varies significantly with the path, and the search for the best option is demanding, path randomization always proves more adequate.

Along with the numerical analysis, we have also proposed an algorithm to search for new DD schemes. This has resulted in an extremely efficient pulse sequence based on frequent path alteration. Interestingly, however, this sequence turned out to be outperformed by a very simple scheme which combined the initial pulses from the algorithm sequence with randomization. The main take-away message is that even though an 
optimal deterministic sequence may always exist for a particular system at a specific final time, identifying it may be beyond reach, 
in which case resorting to simpler, yet efficient randomized sequences becomes a practical method of choice.

At last, the effects of two control non-idealities -- finite width pulses and flip-angle errors -- have been quantified. Deterministic protocols
appear to be better protected against such imperfections, although the relative gain due to randomization still dominates if the errors are relatively small.  A complete analysis of fault-tolerance requires,
however, consideration of additional compensation mechanisms along with
randomization, which we intend to address elsewhere.

\subsection{Outlook}

The selection of an adequate DD protocol ultimately depends on details about the system and the control objective to be achieved. A sequence like {\tt PCDD}$_2$, for example, is excellent to decouple a single qubit from its surrounding bath \cite{Wen01,Wen02,Wen03}, but performs poorly at freezing evolution in a spin chain with $NN$ interactions. Similarly, 
the WAHUHA sequence combined with cyclic permutations lead to third order DD of the dipolar Hamiltonian~\cite{HaeberlenBook}, whereas {\tt PSCPD}$_2$ is unable to cancel ${\bar H}^{(2)}$ in $H_{NN}$. When the control pulses aim at complete refocusing, the total time during which information
needs to be stored is a decisive factor in the choice of a protocol.
By comparing the evolution times in Figs.~\ref{fig:NN8} and \ref{fig:optimal}, for example, one sees that {\tt SRPD} is a good enough method in the first situation, although not worth consideration in the latter. Another important consideration stems from the desired control goal: the removal of unwanted evolution, independent of the choice of initial state, as addressed here by analyzing the decay of entanglement fidelity; or the preservation of a specific, known initial state.
The latter scenario may allow for the development of dedicated pulse sequences ensuring yet better performance -- as exemplified by long-time coherence saturation effects observed in both NMR spin-locking experiments
\cite{HaeberlenBook} and in quantum information storage \cite{Wen01,Wen02,Wen03}.

Throughout this work, the time interval between consecutive pulses 
in {\tt PDD} has a fixed value $\Delta t$, while for other protocols actual rotations may be separated by some integer multiples of 
$\Delta t$. If this constraint is relaxed, so that consecutive rotations may be arbitrarily spaced, substantial freedom is added in principle to DD design.  In this sense, the existence of optimized sequences for specific control settings, as in \cite{Uhrig2007}, clearly points to the potential of unevenly-spaced sequences for higher-order DD. 
The analysis and combination of multiple control time scales and different various angles of rotation is clearly an issue which deserves additional exploration in the context of randomization, along with the identification of a QIP platform which may be suitable to experimentally test some of the benefits predicted for randomized coherent control.

\newpage
\begin{acknowledgments}

It is a pleasure to thank Paola Cappellaro, David Cory, Evan Fortunato, Malcom Levitt, and Chandrasekhar Ramanathan for useful discussions and input at various stages during this work. Partial support from the National Science Foundation through the ``Physics at the Information Frontier" Program, Grant No. PHY-0555417, is gratefully acknowledged.

\end{acknowledgments}

\appendix

\section{Dominant terms in the average Hamiltonian}

Here we consider the general Hamiltonian with $NN$ interactions, ${\tilde H}_{Z+NN}^R$ (\ref{HR-Zbi}), where both anisotropy ($\alpha \neq 1$) as well as one-body terms may be present, and provide explicit results for the first three contributions to the average Hamiltonian in the case of deterministic protocols.

\subsection{Lowest-order average Hamiltonian: ${\bar H}^{(0)}$}

Representations ${\cal G}_{XZ}$ and ${\cal G}_{ZY}$ (\ref{repres}), 
which involve group elements in the $z$ direction and also representations affecting only half of the qubits cannot cancel all one-body terms.  If complete refocusing of the Hamiltonian is the goal, ${\cal G}_{XY}$ is the representation to be used, for it guarantees ${\bar H}^{(0)}(T_c)=0$. Let us consider two particular
pulse sequences characterized by the following paths:
\begin{eqnarray}
&&{\rm Path}\; 1: \hspace{0.3 cm} \{ \mathbb{1},\; X_1 Y_2  \ldots 
X_{N-1} Y_{N} , \; X_1 X_3 \ldots X_{N-1} , \; Y_2 Y_4 \ldots Y_{N}  \},
\nonumber \\
&&{\rm Path}\; 2: \hspace{0.3 cm} \{ \mathbb{1},\; X_1 X_3 \ldots X_{N-1}  ,
\;  X_1 Y_2  \ldots 
X_{N-1} Y_{N}, \; Y_2 Y_4 \ldots Y_{N}\}.
\label{paths}
\end{eqnarray}

\subsection{First-order contribution to the 
average Hamiltonian: ${\bar H}^{(1)}$}

For {\tt PDD} sequences from ${\cal G}_{XY}$ that change the sign of the Ising interaction after every $\Delta t$, such as Path $2$, we find
\begin{equation}
{\bar H}^{(1)} (T_c)= 
\mp J \Delta t \left[
\sum_{i-{\rm odd}}
\frac{(\delta_i + \delta_{i+1})}{2} Y_i X_{i+1}  +
\sum_{i-{\rm even}}
\frac{(\delta_i + \delta_{i+1})}{2} X_i Y_{i+1} 
\right]
\pm J^2 \Delta t
\sum_{i=1}^{N-2} \left( X_i Z_{i+1} Y_{i+2}
+ Y_i Z_{i+1} X_{i+2} \right) , \label{H1Zee}
\end{equation}

\par\noindent
while for Path $1$, Eq.~(\ref{H1Ay}) still holds.
The interplay between anisotropy and qubit frequencies becomes now a determining factor in the selection of an appropriate group path. 

\subsection{Second-order contribution to the 
average Hamiltonian: ${\bar H}^{(2)}$}

We compute ${\bar H}^{(2)}$ for the two pulse sequences
of ${\cal G}_{XY} $ in Eq.~(\ref{paths}). The following results are found for {\tt SDD}, {\tt PCDD}$_2$, and {\tt PSCPD}$_2$. Notice that for reasons explained in Sec.VI.A, ${\bar H}^{(2)}(2T_c)$ for {\tt SDD} equals
${\bar H}^{(2)}(T_c)$ for {\tt PDD}.
\begin{eqnarray}
&& {\underline {\rm Path\; 1}:}
\label{path1}
\\
&& {\tt SDD}: \hspace{0.2 cm} {\bar H}^{(2)} =
-\frac{A}{3} 
 L_a
-2JA
\left\{ D_z +Q_z
-\alpha \left(Y_1 Y_2 + Y_{N-1} Y_N +2 \sum_{i=2}^{N-2} 
Y_i Y_{i+1} \right)
 +2\alpha \sum_{i=1}^{N-3} Z_i X_{i+1} X_{i+2} Z_{i+3}
\right\}
\nonumber
\\
&& {\tt PCDD}_2: \hspace{0.2 cm} 
\hspace{0.2 cm} {\bar H}^{(2)} =
- \frac{A}{3} L_a
-2JA
( D_z +Q_z )
\nonumber
\\
&& {\tt PSCPD}_2: \hspace{0.2 cm} {\bar H}^{(2)} =
\frac{A}{6} L_b
 + JA( D_x + Q_x) 
\nonumber
\end{eqnarray}

\begin{eqnarray}
&& {\underline {\rm Path\; 2}:} 
\label{path2}
\\
&& {\tt SDD}: \hspace{0.2 cm} {\bar H}^{(2)} =
(\Delta t)^2 \left\{ (\delta_1 + \delta_2) Z_1 + 
\sum_{i-{\rm odd}} (\delta_{i-1} + 2\delta_{i} + \delta_{i+1}) Z_i
+ (\delta_{N-1} + \delta_N) Z_{N-{\rm odd}}
\right\} \hspace{3 cm}
\nonumber
\\
&& + \frac{J}{2} (\Delta t)^2 
\left[ \sum_{i-{\rm odd}} (\delta_{i} + \delta_{i+1})\delta_{i+1}
Y_i Y_{i+1} +
\sum_{i-{\rm even}} (\delta_{i} + \delta_{i+1})\delta_{i}
Y_i Y_{i+1}
\right]
\nonumber
\\
&& - 
 \frac{A}{3} \left\{L_b + \frac{3}{\alpha}
\sum_{i-{\rm even}}
\left[
-2 (\delta_i + \delta_{i+1} + \delta_{i+2}) X_i Z_{i+1}X_{i+2} + 
(\delta_i + \delta_{i+2}) Y_i Z_{i+1} Y_{i+2}
\right]
\right\}
\nonumber
\\
&& - 2JA
\left\{ D_x +Q_x
-\frac{1}{\alpha} \left( Y_1 Y_2 + Y_{N-1} Y_N +2 \sum_{i=2}^{N-2} Y_i Y_{i+1}
\right)
+\frac{2}{\alpha} \sum_{i=1}^{N-3} X_i Z_{i+1} Z_{i+2} X_{i+3}
\right\}
\nonumber
\\
&&{\tt PCDD}_2: \hspace{0.2 cm} {\bar H}^{(2)} =
-  \frac{A}{3} L_b  - 2JA (D_x +Q_x)
\nonumber
\\
&&{\tt PSCPD}_2: \hspace{0.2 cm} {\bar H}^{(2)} = 
 \frac{A}{6} L_a 
 +J A (D_z + Q_z)
\nonumber
\end{eqnarray}
where the following quantities have been introduced:
\begin{eqnarray}
A&=&J^2  (\Delta t)^2 \alpha , 
\nonumber
\\
L_a&=&\sum_{i-{\rm odd}} \left[(\delta_i - \delta_{i+1})
Y_i Y_{i+1} Z_{i+2} 
-(\delta_{i+1} - \delta_{i+2})
Z_i Y_{i+1} Y_{i+2}\right],
\nonumber
\\
&+&\sum_{i-{\rm even}} \left[(\delta_i - \delta_{i+1})
X_i X_{i+1} Z_{i+2} 
- (\delta_{i+1} - \delta_{i+2})
Z_i X_{i+1} X_{i+2}
\right],
\nonumber
\\
L_b&=&
- \sum_{i-{\rm odd}} \left[(2\delta_i + \delta_{i+1})
Y_i Y_{i+1} Z_{i+2} 
+(\delta_{i+1} + 2 \delta_{i+2})
Z_i Y_{i+1} Y_{i+2}\right],
\nonumber
\\
&& +
\sum_{i-{\rm even}} \left[(\delta_i +2 \delta_{i+1})
X_i X_{i+1} Z_{i+2} 
+ (2\delta_{i+1} + \delta_{i+2})
Z_i X_{i+1} X_{i+2}
\right],
\nonumber
\\
D_z&=&\frac{2}{3} \sum_{i=1}^{N-2}(X_{i} X_{i+2} + Y_i Y_{i+2} - 
2 Z_i Z_{i+2}) 
\nonumber
\\
D_x&=& \frac{2}{3} \sum_{i=1}^{N-2}(  Y_i Y_{i+2} + 
Z_i Z_{i+2} - 2 X_{i} X_{i+2}) ,
\nonumber
\\
Q_z&=&
\frac{1}{3} \sum_{i=1}^{N-3} \left[ 
X_i X_{i+2} \left( 2 Y_{i+1} Y_{i+3} - Z_{i+1} Z_{i+3} \right) + 
Y_i Y_{i+2} \left( 2 X_{i+1} X_{i+3} - Z_{i+1} Z_{i+3} \right) 
- Z_i Z_{i+2} \left(  X_{i+1} X_{i+3} + Y_{i+1} Y_{i+3} \right)
\right],
\nonumber
\\
Q_x&=&
\frac{1}{3} \sum_{i=1}^{N-3} \left[ 
Y_i Y_{i+2} \left( 2 Z_{i+1} Z_{i+3} - X_{i+1} X_{i+3} \right) 
+ Z_i Z_{i+2} \left( 2 Y_{i+1} Y_{i+3} - X_{i+1} X_{i+3} \right)
-X_i X_{i+2} \left( Y_{i+1} Y_{i+3} + Z_{i+1} Z_{i+3} \right)
\right] .
\nonumber
\end{eqnarray}


\begin{thebibliography}{86}
\bibitem[{Pry()}]{Pryadko_private}
\bibinfo{note}{The leading term results for an isolated qubit under the {\tt
  CDD} protocol were first indicated to us by Leonid Pryadko (private
  communication).}
\end{thebibliography}
\end{document}